%% file: article.tex
\documentclass[a4paper,twosided]{article}
\pdfoutput=1
\usepackage{amsfonts,amstext,amsmath,amssymb,stmaryrd,mathrsfs,calc,amsthm,xspace,mathdots,mathtools,bm}
\DeclareMathAlphabet{\mathpzc}{OT1}{pzc}{m}{it}
\usepackage{array,multirow}
\makeatletter

\newcommand{\ctrl}[1][\@empty]{\ifx#1\@empty{,f}\else{,#1}\fi}
\renewcommand\section{\@startsection {section}{1}{\z@}%
                                   {-3.5ex \@plus -1ex \@minus -.2ex}%
                                   {2.3ex \@plus.2ex}%
                                   {\normalfont\Large\bfseries\boldmath}}
\renewcommand\subsection{\@startsection{subsection}{2}{\z@}%
                                     {-3.25ex\@plus -1ex \@minus -.2ex}%
                                     {1.5ex \@plus .2ex}%
                                     {\normalfont\large\bfseries\boldmath}}
\renewcommand\subsubsection{\@startsection{subsubsection}{3}{\z@}%
                                     {-3.25ex\@plus -1ex \@minus -.2ex}%
                                     {1.5ex \@plus .2ex}%
                                     {\normalfont\normalsize\bfseries\boldmath}}
\renewcommand\paragraph{\@startsection{paragraph}{4}{\z@}%
                                    {3.25ex \@plus1ex \@minus.2ex}%
                                    {-1em}%
                                    {\normalfont\normalsize\bfseries}}
\renewcommand\subparagraph{\@startsection{subparagraph}{5}{\parindent}%
                                       {3.25ex \@plus1ex \@minus .2ex}%
                                       {-1em}%
                                      {\normalfont\normalsize\bfseries\boldmath}}
\makeatother
\newcommand{\FGH}[1]{\mathfrak{F}_{#1}} %
\newcommand{\bad}{L}   %
\newcommand{\ubound}{M}%
\newcommand{\uboundd}[3][f]{\ubound_{\ifx#31{}\else{{#3}\times}\fi\{#2\}\ctrl[#1]}}
\newcommand{\urec}[2][f]{G_{{#2}\ctrl[#1]}}
\newcommand{\llex}{\ell}
\newcommand{\llexd}[2][f]{\llex_{{#2}\ctrl[#1]}}
\newcommand{\lex}{\leq_{\mathrm{lex}}}

\newcommand{\Mstep}{\rightarrow_\text{M}}
\newcommand{\Istep}{\rightarrow_\text{I}}
\newcommand{\step}[1]{\xrightarrow{\!\!#1\!\!}}
\newcommand{\Post}{\mathit{Post}}
\newcommand{\Reach}{\mathit{Reach}}
\newcommand{\Conf}{\mathit{Conf}\!}
\newcommand{\lbound}{\llexd} %
\newcommand{\badd}[3][f]{\bad_{\ifx#31{}\else{{#3}\times}\fi\{#2\}\ctrl[#1]}} %
\newcommand{\Lbound}[2][f]{o_{{#2}\ctrl[#1]}} %
\newcommand{\eqdef}                     %
  {\stackrel{\scriptscriptstyle \mathrm{def}}{=}}
\newcommand{\equivdef}                  %
  {\stackrel{\scriptscriptstyle \mathrm{def}}{\Leftrightarrow}}
\newcommand{\tetra}[2]{\underbrace{{#1}^{{#1}^{\iddots\raisebox{1ex}{{\tiny
            {#1}}}}}}_{#2\text{ times}}}

\newcommand{\len}[1]{|#1|}
\newcommand{\leni}[1]{\len{#1}_\infty}
\newcommand{\tup}[1]{\langle #1 \rangle}
\newcommand{\N}{\mathbb{N}}
\newcommand{\Z}{\mathbb{Z}}

\newcommand{\cutsub}{\mathrel{\mathop{-}^{\hspace{-1.15ex}.}\,}}
\usepackage{thmtools,thm-restate}
\declaretheorem[numberwithin=section]{theorem}
\declaretheorem[sibling=theorem]{lemma}

\declaretheorem[sibling=theorem]{corollary}
\declaretheorem[sibling=theorem]{fact}
\declaretheorem[sibling=theorem,style=definition]{question}

\declaretheorem[sibling=theorem,style=definition]{example}
\declaretheorem[parent=theorem,style=remark]{claim}
\declaretheorem[sibling=theorem,style=remark]{remark}
\makeatletter
\def\ps@firstpage{\let\@mkboth\@gobbletwo%
\def\@oddhead{}%
\def\@oddfoot{%
\hbox{%
  \vbox to 100 pt{\footnotesize%
    \begin{minipage}[t]{\textwidth}\vspace*{2em}\par Published as Figueira, D., Figueira, S., Schmitz, S., and Schnoebelen, {\relax Ph}., 2011.
\newblock {A}ckermannian and primitive-recursive bounds with {D}ickson's
  {L}emma.
\newblock In \emph{LICS 2011}\natconfdetails[\emph{26th}]{\emph{Annual IEEE
  Symposium on Logic in Computer Science}}, pages 269--278. IEEE.
\newblock \href{http://dx.doi.org/10.1109/LICS.2011.39}
  {\nolinkurl{doi:10.1109/LICS.2011.39}}..\end{minipage}}\hfill
    }}
\def\@evenhead{}
\def\@evenfoot{}}%
\makeatother
\usepackage{natbib}
\newcommand{\natconfdetails}[2][]{, #1 #2}
\usepackage{hyperref}

\providecommand{\urlstyle}[1]{}
\providecommand{\doi}[1]{\href{http://dx.doi.org/#1}{\nolinkurl{doi:#1}}}

\begin{document}
\input{version}

\title{Ackermannian and Primitive-Recursive Bounds with
  Dickson's Lemma\thanks{Work supported by the Agence Nationale de la Recherche, grant ANR-06-SETIN-001, and by the Future and Emerging Technologies (FET) programme
within the Seventh Framework Programme for Research of the European Commission, under the FET-Open
grant agreement FOX, number FP7-ICT-233599.}}
\author{Diego Figueira$^1$ \and Santiago Figueira$^2$ \and Sylvain Schmitz$^3$
  \and Philippe Schnoebelen$^3$}
\date{{\small 
      $^1$ University of Edinburgh\\
      \href{mailto:dfigueir@inf.ed.ac.uk}{dfigueir@inf.ed.ac.uk}\\
      $^2$ Dept.\ of Computer Science, FCEyN, University of Buenos
      Aires \& CONICET\\
      \href{mailto:santiago@dc.uba.ar}{santiago@dc.uba.ar}\\
      $^3$ LSV, ENS Cachan \& CNRS\\
      \href{mailto:sylvain.schmitz@lsv.ens-cachan.fr,phs@lsv.ens-cachan.fr}{\{schmitz,phs\}@lsv.ens-cachan.fr}\\
}}
\maketitle
\begin{abstract}
Dickson's Lemma is a simple yet powerful tool widely used in decidability
proofs, especially when dealing with counters or related data structures in
algorithmics, verification and model-checking, constraint solving, logic,
etc. While Dickson's Lemma is well-known, most computer scientists are not
aware of the complexity upper bounds that are entailed by its use. This is
mainly because, on this issue, the existing literature is not very
accessible.

We propose a new analysis of the length of \emph{bad sequences} over
$(\N^k,\leq)$, improving on earlier results and providing upper bounds that
are essentially tight. This analysis is complemented by a ``user guide''
explaining through practical examples how to easily derive complexity upper
bounds from Dickson's Lemma.
\end{abstract}
\thispagestyle{firstpage}

\input{intro}

\input{polwqo}

\input{dickson}

\input{r-bad}

\input{bounding}

\subsection{Classifying $\ubound$ in the Fast Growing Hierarchy}\label{sec-fgh-ubound}
\input{hierarchy-upper}

\section{Lower Bound}\label{sec:lbound}
\input{hierarchy-lower}

\section{Applications}\label{sec-applications}
\input{applications}
\section{Related Work}\label{sec:related}

\input{related}

\input{conclusion}

\bibliography{journals,mcaloon}
\clearpage
\appendix
\section{Proofs Omitted from the Main Text}
\input{app-decomposition}
\input{hierarchy-computations}
\end{document}

%% file: version.tex
\def\svnversion{%
\unskip, 
July 19, 2011
}

%% file: intro.tex
\newcommand{\Nat}{{\mathbb{N}}}%
\newcommand{\xxx}{{\mathbf{x}}}%
\newcommand{\yyy}{{\mathbf{y}}}%
\newcommand{\zzz}{{\mathbf{z}}}%
\newcommand{\vvv}{{\mathbf{v}}}%
\newcommand{\egdef}{\eqdef}

\section{Introduction}
For some dimension $k$, let $(\N^k,\leq)$ be the set of $k$-tuples of
natural numbers ordered with the natural product ordering
\begin{multline*}
x=\tup{x[1],\dots,x[k]}\;\leq\;y=\tup{y[1],\dots,y[k]}
\equivdef
x[1]\leq y[1] \wedge \cdots \wedge x[k]\leq y[k]\;.
\end{multline*}
Dickson's Lemma is the statement that $(\N^k,\leq)$ is a
well-quasi-ordering (a ``wqo''). This means that there exist no infinite
strictly decreasing sequences $x_0>x_1>x_2>\cdots$ of $k$-tuples, and that
there are no infinite antichains, i.e., sequences of pairwise incomparable
$k$-tuples~\citep{kruskal72,milner85}.  Equivalently, every infinite
sequence $\xxx=x_0,x_1,x_2,\dots$ over $\N^k$ contains an \emph{increasing
  pair} $x_{i_1}\leq x_{i_2}$ for some $i_1<i_2$%
.  We say that sequences with an increasing pair
$x_{i_1}\leq x_{i_2}$ are \emph{good} sequences%
. We say that a
sequence that is not %
good is %
\emph{bad}. Dickson's Lemma states that every
infinite sequence over $\N^k$ is %
good%
, i.e., that %
bad
sequences are finite.

\paragraph{Using Dickson's Lemma}
``The most frequently rediscovered mathematical theorem'' according
to \citep[p.~184]{grobner}, Dickson's Lemma plays a fundamental role
in several areas of computer science, where it is used to prove that
some algorithmic constructions terminate,  that some sets are
finite, or semilinear, \textit{etc}.  In
\autoref{sec-applications}, we give examples
dealing with counter machines and Petri nets because we are more
familiar with this area, but many others exist.
\begin{example}[label=ex:choice]
The following simple program is shown in \citep{PodelskiR-lics04} to
terminate for every input $\tup{a,b}\in\N^2$:\\[.5em]
{\small\begin{tabular}{l}
 \textsc{choice} $(a,b)$\\
 \textbf{while} {$a>0\wedge b>1$}\\
 \qquad $\tup{a,b}\longleftarrow\tup{a-1,a}$\\
 \quad\textbf{or}\\
 \qquad $\tup{a,b}\longleftarrow\tup{b-2,a+1}$\\
 \textbf{end}
\end{tabular}}\\[.5em]
We leave it to the reader to check that, in fact,
any sequence of successive configurations
$x_0=\tup{a,b},x_1,x_2,\ldots$ of this program is a bad sequence over
$\N^2$, and is thus finite by Dickson's Lemma.  Let
\textsc{Time}$(a,b)$ be the maximal number of times the \textbf{while}
loop of \textsc{choice} can be executed---a natural complexity measure. If we could bound the length
of bad sequences over $\N^2$ that start with $\tup{a,b}$, then we
would have an upper-bound on \textsc{Time}$(a,b)$.\qed
\end{example}

In order to bound the running time of algorithms that
rely on Dickson's Lemma, it is usually necessary to know (or to bound)
the value of the index $i_2$ in the first increasing pair $x_{i_1}\leq x_{i_2}$.
It is widely felt, at least in the field of verification and
model-checking, that relying on Dickson's Lemma when proving
decidability or finiteness does not give any useful information
regarding complexity, or that it gives upper bounds that are not
explicit and/or not meaningful.  Indeed, bad sequences can be
arbitrarily long.

\paragraph{The Length of Bad Sequences}
It is easy to construct arbitrarily long bad sequences, even when starting from
a fixed first element. Consider $\N^2$ and fix $x_0=\tup{0,1}$. Then the following
\[
\tup{0,1}, \, \tup{L,0}, \, \tup{L-1,0}, \, \tup{L-2,0}, \, \cdots\, \tup{2,0}, \, \tup{1,0} 
\]
is a bad sequence of length $L+1$. What makes such examples possible is the
``uncontrolled'' jump from an element like $x_0$ to an \emph{arbitrarily} large
next element like here $x_1=\tup{L,0}$. Indeed, when one only considers bad
sequences displaying some controlled behaviour (in essence, bad sequences
of bounded complexity), upper bounds on their lengths certainly exist.

Let us fix a \emph{control function} $f:\N\to \N$. We say that a sequence
$\xxx=x_0,x_1,\dots$ over $\N^k$ is \emph{$t$-controlled} for some $t$
in $\N$ %
if the infinity norm of the $x_i$ verifies $\len{x_i}_\infty < f(i+t)$
for all indexes $i=0,1,\ldots$  Then, for fixed $k$, $t$, %
and $f$,
there are only finitely many $t$-controlled %
bad sequences (by
Dickson's Lemma \textit{cum} K\H{o}nig's Lemma) and a maximum length
exists. This maximum length %
can even be
computed if $f$ is recursive.  %

In this paper, we write $\bad_{k,f}(t)$ for the maximal length of a
$t$-controlled bad sequence (given $f$, and a dimension $k$) and bound
it from above via a new decomposition approach. These results are
especially useful when we study $\bad_{k,f}(t)$ as a function of $t$,
i.e.\ when we prove that the function $\bad_{k,f}$ is majorized by a
function in a given complexity class. The literature already contains
upper bounds on $\bad_{k,f}$ (see~\autoref{sec:related}) but these
results are not widely known. Most prominently, \citet{mcaloon} shows
that for linear $f$, $\bad_{k,f}$ is primitive-recursive for each
fixed $k$, but is not primitive-recursive when $k$ is not fixed. More
precisely, for every $k$, $\bad_{k,f}$ is at level $\FGH{k+1}$ of the
Fast Growing Hierarchy.\footnote{In truth, \citeauthor{mcaloon} is not
  that explicit.  The $\FGH{k+1}$ upper bound is extracted from his
  construction by \citet{clote}, who also proposed a simple derivation
  for an upper bound at level
  $\FGH{k+6}$.}%
To quote \citet{clote}, ``This suggests the
question whether $\FGH{k+1}$ is the best possible.''

\paragraph{Our Contribution}
We present a self-contained and elementary proof,
markedly simpler and more general than \citeauthor{mcaloon}'s, but yielding an improved
upper bound: for linear control functions, $\bad_{k,f}$ is at level
$\FGH{k}$, and more generally, for a control function $f$ in
$\FGH{\gamma}$%
, $\bad_{%
  k,f}$ is at level
$\FGH{\gamma+k-1}$.
\begin{example}[continues=ex:choice]
  Setting $f(x)=x+1$ makes every sequence of configurations of
  \textsc{choice}$\,(a,b)$ a $(\max(a,b))$-controlled bad sequence,
  for which our results incur an elementary length in $\FGH{2}$ as a
  function of $\max(a,b)$.
\qed\end{example}
That ``\textsc{Time}$(a,b)$ is in $\FGH{2}$'' is a very coarse bound, but as we will
see in \autoref{sec:lbound}, allowing larger dimensions or more
complex operations quickly yield huge complexities on very simple
programs similar to \textsc{choice}.  In fact, we also answer
\citeauthor{clote}'s question, and show that our upper bounds are
optimal. %

More precisely, our main technical contributions are %
\begin{itemize}
\item We substantially simplify the problem by considering a richer
 setting for our analysis: all disjoint unions of powers of $\N$.  This lets us provide
  finer and simpler decompositions of bad sequences
  (\autoref{sec-dickson}),
  from which one extracts upper bounds on their lengths (\autoref{sec-max-ubound}).
\item  We completely separate the decomposition issue
 (from complex to simple wqo's, where $f$ is mostly irrelevant)
 from the question of locating the bounding function in the Fast Growing
 Hierarchy (where $f$ becomes relevant); see  \autoref{sec-fgh-ubound}.
\item We obtain new bounds that are essentially tight in terms of the Fast Growing
  Hierarchy; see \autoref{sec:lbound}. Furthermore, these
  bounds are tight even when considering the coarser lexicographic ordering%
  .
\item We describe another benefit of our setting: it
  accommodates in a smooth and easy way an extended notion of bad sequences
  where the length of the forbidden increasing subsequences is a parameter
  (\autoref{sec-long-r-bad}).
\end{itemize}
In addition we provide (in \autoref{sec-applications}) a few examples
showing how to use bounds on $\bad_{%
  k,f}$ in practice.  This section
is intended as a short ``user guide'' showing via concrete examples how to
apply our main result and derive upper bounds from one's use of Dickson's Lemma.
We do not claim that we show new results for these examples, although the
existence of the
bounds we obtain is hardly known at all.  %
The examples we picked are
some of our favorites (many others exist, see
\autoref{sec:related} for a few references). In particular, they involve
algorithms or proofs that do not directly
deal with bad sequences over $(\N^k,\leq)$:
\begin{itemize}
\item  
   programs shown to terminate using \emph{disjunctive termination
     arguments} (\autoref{sub:rank}),
\item
   emptiness for \emph{increasing counter automata} with applications to
   questions for \textsc{XPath} fragments on data words
   (\autoref{sub:xpath}), %
and
\item 
   \emph{\citeauthor{kmtree} coverability trees} and their applications, %
 (\autoref{sub:km}).
\end{itemize}

%% file: polwqo.tex
\section{WQO's Based on Natural Numbers}
\label{sec-polwqo}

The disjoint union, or ``sum'' for short, of two sets $A$ and $B$ is
denoted $A+B$, the sum of an $I$-indexed family $(A_i)_{i\in I}$ of sets is
denoted $\sum_{i\in I} A_i$. While $A+B$ and $\sum_i A_i$ can be seen as,
respectively, $A\times\{1\}\cup B\times\{2\}$ and $\bigcup_i
A_i\times\{i\}$, we abuse notation and write $x$ when speaking of an
element $(x,i)$ of $\sum_i A_i$.

Assume $(A_1,\leq_1)$ and $(A_2,\leq_2)$ are ordered sets. The product
$A_1\times A_2$ is equipped with the usual product ordering: $(x,y)\leq
(x',y')\equivdef x\leq_1 x' \wedge y\leq_2 y'$. The sum $A_1+A_2$ is
equipped with the usual sum ordering given by
\begin{gather*}
                x\leq x'   
\equivdef   
             \bigl( x,x'\in A_1  \wedge x\leq_1 x' \bigr) 
        \vee \bigl( x,x'\in A_2  \wedge x\leq_2 x' \bigr)\,.
\end{gather*}
It is easy to see that $(A_1\times A_2,\leq)$ and $(A_1+A_2,\leq)$ are wqo's
when $(A_1,\leq_1)$ and $(A_2,\leq_2)$ are. This immediately extends to
$\prod_{i\in I}A_i$ and $\sum_{i\in I}A_i$ when the index set $I$ is
finite. Note that this allows inferring that $(\N^k,\leq)$ is a wqo
(Dickson's Lemma) from the fact that $(\N,\leq)$ is.

A key ingredient of this paper is that we consider finite sums of finite
powers of $\N$, i.e., sets like, e.g., $2\times\N^3+\N$ (or
equivalently $\N^3+\N^3+\N^1$, and more generally of the form
$\sum_{i\in I}\N^{k_i}$). With $S=\sum_{i\in I}\N^{k_i}$, we associate
its \emph{type} $\tau$, defined as the multiset $\{k_i~|~i\in I\}$, and let
$\N^\tau$ denote $S$ (hence $\N^{\{k\}}$ is $\N^k$ and
$\N^\emptyset$ is $\emptyset$).

Types such as $\tau$ can be seen from different angles. The multiset point
of view has its uses, e.g., when we observe that
$\N^{\tau_1}+\N^{\tau_2}=\N^{\tau_1+\tau_2}$. But types can also be
seen as functions $\tau:\N\rightarrow\N$ that associate with each power
$k\in\N$ its multiplicity $\tau(k)$ in $\tau$. We define the sum
$\tau_1+\tau_2$ of two types with
$(\tau_1+\tau_2)(k)\egdef\tau_1(k)+\tau_2(k)$ and its multiple
$p\times\tau$, for $p\in\N$, by $(p\times\tau)(k)\egdef p.\tau(k)$. As
expected, $\tau-\tau_1$ is only defined when $\tau$ can be written as some
$\tau_1+\tau_2$, and then one has $\tau-\tau_1=\tau_2$.
\medskip

There are two natural ways of comparing types: the inclusion ordering
\begin{gather}
\tau_1\subseteq\tau_2 \equivdef \exists \tau':\tau_2=\tau_1+\tau'
\end{gather}
and the multiset ordering defined by transitivity and
\begin{alignat}{2}
\label{eq-<m-1}
      \tau  &<_m \{k\}	     &&\equivdef \text{ $k>l$ for all $l\in\tau$,}
\\
\label{eq-<m-2}
\tau_1+\tau & <_m \tau_2+\tau && \equivdef \tau_1<_m\tau_2\:.
\end{alignat}
Note how Eq.~\eqref{eq-<m-1} entails $\emptyset<_m\{k\}$.
Then Eq.~\eqref{eq-<m-2} further yields $\emptyset\leq_m\tau$ for any
$\tau$ (using transitivity).
In fact, the multiset ordering is a well-founded linear extension of
the inclusion ordering \citep[see][]{dershowitz79}. This is the ordering we
use when we reason ``by induction over types''.

%% file: dickson.tex
\section{Long Bad Sequences over $\N^{\tau}$}
\label{sec-dickson}

Assume a fixed, increasing, control function $f:\N\to\N$ with
$f(0)>0$; we keep $f$ implicit to simplify notations, until
\autoref{sec-fgh-ubound} where the choice of control function will
become important.  For $t\in\N$, we say that a sequence
$x_0,x_1,\dots,x_l$ over $\N^\tau$ is \emph{$t$-controlled} if
$\len{x_i}_\infty<f(i+t)$ for all $i=0,1,\ldots,l$, where
$\len{x_i}_\infty\egdef\max\{ x_i[j]~|~j=1,\ldots,\mathit{dim}(x_i)\}$
is the usual infinity norm.
Let $\bad_\tau(t)$ be the length of the longest $t$-controlled bad
sequence over $\N^\tau$.

In simple cases, $\bad_\tau(t)$ can be evaluated exactly.  For example
consider $\tau=\{0\}$. Here $\N^\tau$, i.e., $\N^0$, only contains one element,
the empty tuple $\tup{}$, whose norm is $0$, so that every sequence over
$\N^\tau$ is $t$-controlled because $f(0)>0$, and is good as soon as
its length is greater than or equal to $2$. Hence
\begin{align}
\label{eq-L0}
			       \bad_{\{0\}}(t)&=1\;,
\shortintertext{and more generally for all $r\geq 1$}
\label{eq-Ln0}
			   \bad_{r\times\{0\}}(t)&=r\;.
\end{align}
Note that this entails $\bad_\emptyset(t) = \bad_{0\times\{0\}}(t) = 0$ as
expected: the only sequence over $\N^\emptyset$ is the empty sequence.

The case $\tau=\{1\}$ is a little bit more interesting. A bad sequence
$x_0,x_1,\dots,x_l$ over $\N^{\{1\}}$, i.e., over $\N$, is a decreasing
sequence $x_0>x_1>\cdots>x_l$ of natural numbers. Assuming that the
sequence is $t$-controlled means that $x_0<f(t)$. (It is further
required that $x_i<f(t+i)$ for every $i=1,\dots,l$ but here this brings no
additional constraints since $f$ is increasing and the sequence must be
decreasing.)  It is plain that $\bad_{\{1\}}(t)\leq f(t)$, and in fact
\begin{gather}
\label{eq-L1}
			      \bad_{\{1\}}(t)=f(t)
\end{gather}
since the longest $t$-controlled bad sequence is exactly
\[
		     f(t)-1,\, f(t)-2,\, \ldots,\, 1,\, 0\;.
\]

\paragraph{Decomposing Bad Sequences over {$\N^\tau$}}
After these initial considerations, we turn to the general case. It is
harder to find \emph{exact} formulae for $\bad_\tau(t)$ that work generally.
In this section, we develop inequations providing upper bounds for
$\bad_\tau(t)$ by induction over the structure of $\tau$. These inequations
are enough to prove our main theorem.

Assume $\tau=\{k\}$ and consider a $t$-controlled bad sequence
$\xxx=x_0,x_1,\dots,x_l$ over $\N^k$. Since $\xxx$ is $t$-controlled, $x_0$
is bounded and $x_0\leq\tup{f(t)-1,\ldots,f(t)-1}$. Now, since $\xxx$ is
bad, every $x_i$ for $i>0$ must have $x_i[j]<x_0[j]$ for at least one $j$
in $1,\ldots,k$. In other words, every element of the \emph{suffix
 sequence} $x_1,\dots, x_l$ belongs to at least one region
\begin{equation*}
		      R_{j,s}=\{x\in\N^k\mid x[j]=s\}
\end{equation*}
for some $1\leq j\leq k$ and $0\leq s < f(t)-1$. The number of regions is
\begin{equation}
\label{eq-def-Nkt}
  N_{k}(t)\egdef k\cdot (f(t)-1)\;.
\end{equation}
By putting every $x_i$ in
one of the 
regions, we decompose the  suffix sequence into $N_{k}(t)$ subsequences,
some of which may be empty.

We illustrate this with an example. Let $k=2$ and consider the following bad
sequence over $\N^2$
{\small\begin{equation*}
\label{eq-D0}
\xxx =
\tup{2,2},\,
\tup{1,5},\,
\tup{4,0},\,
\tup{1,1},\,
\tup{0,100},\,
\tup{0,99},\,
\tup{3,0}\;.
\end{equation*}}
The relevant regions are $R_{1,0}$, $R_{1,1}$,
$R_{2,0}$, and $R_{2,1}$.  We can put
$x_3=\tup{1,1}$ in either $R_{1,1}$ or $R_{2,1}$, but we have no
choice for the other $x_j$'s. Let us put $x_3$
in $R_{1,1}$; we obtain the following decomposition:
{\small\begin{equation*}
\label{eq-D1}
\tup{2,2},\!
\left[
\setlength{\arraycolsep}{.2pt}
\begin{array}{ccccccr}
 .&
 .&
 .&
\tup{0,100}, &
\tup{0,99}, &
 .&
\: (R_{1,0}: x[1]=0)
\\
 \tup{1,5}, &
 .&
\tup{1,1}, &
 .&
 .&
 .&
\: (R_{1,1}: x[1]=1)
\\
 .&
\tup{4,0}, &
 .&
 .&
 .&
\tup{3,0} &
\: (R_{2,0}: x[2]=0)
\\
 .&
 .&
 .&
 .&
 .&
 .&
\: (R_{2,1}: x[2]=1)
\end{array}
\right]
\end{equation*}}
We have 4 subsequences, one per line.  Each subsequence is bad
(one is even empty). They are not $(t+1)$-controlled if we see them as
\emph{independent} sequences.  For instance, the first subsequence,
``$\tup{0,100},\tup{0,99}$'', is only controlled if $100<f(t+1)$, while in
the original sequence it was only required that $100<f(t+4)$. But they are
$(t+1)$-controlled if we see them as a sequence over the sum type
$4\times\N^2$.

For the next step, we observe that every subsequence has all its elements
sharing a same $x[j]=s$. By disregarding this fixed component, every
subsequence can be seen as a bad sequence over $\N^{k-1}$. In our
example, we get the following decomposition
{\small\begin{equation*}
\tup{2,2},\!
\left[
\setlength{\arraycolsep}{.2pt}
\begin{array}{ccccccr}
 .&
 .&
 .&
\tup{\ast,100}, &
\tup{\ast,99}, &
 .&
\: (R_{1,0}: x[1]=0)
\\
 \tup{\ast,5}, &
 .&
\tup{\ast,1}, &
 .&
 .&
 .&
\: (R_{1,1}: x[1]=1)
\\
 .&
\tup{4,\ast}, &
 .&
 .&
 .&
\tup{3,\ast} &
\: (R_{2,0}: x[2]=0)
\\
 .&
 .&
 .&
 .&
 .&
 .&
\: (R_{2,1}: x[2]=1)
\end{array}
\right]
\end{equation*}}
This way, the suffix sequence $x_1,\dots,x_l$ is seen as a bad sequence
over $\N^{\tau'}$ for $\tau'\egdef N_{k}(t)\times\{k-1\}$. Note that the
decomposition of the suffix sequence always produces a bad,
$(t+1)$-controlled sequence over $\N^{\tau'}$. Hence we conclude that
\begin{gather}
\label{eq-Lk}
	       \bad_{\{k\}}(t)\leq 1+\bad_{N_{k}(t)\times\{k-1\}}(t+1)\;.
\end{gather}
Observe that Eq.~\eqref{eq-Lk} applies even when $k=1$, giving
\begin{align}
  \bad_{\{1\}}(t) & \leq 1+\bad_{(f(t)-1)\times\{0\}}(t+1)
  \notag
  \\
  & = 1 + f(t)-1  =	 f(t)\;.
  \tag{by Eq.~\eqref{eq-Ln0}}
\end{align}
Eq.~\eqref{eq-Lk} still applies in the degenerate ``$k=0$'' case:
here $N_{k}(t)=0$ and the meaningless type ``$\{-1\}$'' is made
irrelevant.

\begin{remark}
  When $k\geq 2$, the inequality in Eq.~\eqref{eq-Lk} cannot be turned into an equality.
  Indeed, a bad sequence over $N_{k}(t)\times\N^{k-1}$ cannot always be
  merged into a bad sequence over $\N^{k}$. As a generic example, take a
  bad sequence $\xxx$ of maximal length over $\N^k$. This sequence ends
  with $\tup{0,...,0}$ (or is not maximal). If we now append another copy
  of $\tup{0,...,0}$ at the end of $\xxx$, the sequence is not bad anymore.
  However, when $k\geq 2$ we can decompose its suffix as a bad sequence
  over $N_{k}(t)\times\N^{k-1}$ by putting the two final $\tup{0,...,0}$'s in the
  different regions $R_{1,0}$ and $R_{2,0}$.
\qed
\end{remark}

The above reasoning, decomposing a sequence over $\N^k$ into a first
element and a suffix sequence over $\N^{\tau'}$ for
$\tau'=N_{k}(t)\times\{k-1\}$, applies more generally for decomposing
a sequence over an arbitrary $\N^\tau$.  Assume $\tau\neq\emptyset$, and let
$\xxx=x_0,x_1,\ldots,x_l$ be a bad sequence over $\N^\tau$. The
initial element $x_0$ of $\xxx$ belongs to $\N^k$ for some $k\in\tau$
and as above $\xxx$ can be seen as $x_0$ followed by a bad subsequence
over $\tau'=N_{k}(t)\times\{k-1\}$, hence the suffix of $\xxx$
can be seen as a bad subsequence over $\tau'+(\tau-\{k\})$.
This calls for special notations: for $k$ in $\tau$ and $t$ in
$\N$, we let 
\begin{equation}
\label{eq-def-taukt}
\tau_{\tup{k,t}} \egdef \tau-\{k\}+N_{k}(t)\times\{k-1\}\;,
\end{equation}
where, for $k=0$, $\tau_{\tup{0,t}}$ is simply $\tau-\{0\}$ since
$N_{0}(t)=0$.

We can now write down the main consequence of our decomposition:
\begin{restatable}{theorem}{Lupperbound}
\label{theo-L-upperbound}
For any $\tau$
\begin{equation*}
\label{eq-L-upperbound}
\bad_{\tau}(t)\leq
\max_{k\in\tau}\left\{1+\bad_{\tau_{\tup{k,t}}}(t+1)\right\}.
\end{equation*}
\end{restatable}

The ``$\max$'' in \autoref{theo-L-upperbound} accounts for allowing a
sequence over $\N^\tau$ to begin with a tuple $x_0$ from any $\N^k$ for
$k\in\tau$. As usual, we let $\max\emptyset\egdef 0$. Note that this
entails $L_\emptyset(t) = 0$, agreeing with \autoref{eq-Ln0}.

%% file: r-bad.tex
\section{Long {$r$}-Bad Sequences}
\label{sec-long-r-bad}

We say that sequences with an increasing subsequence $x_{i_1}\leq
x_{i_2}\leq \cdots\leq x_{i_{r+1}}$ of length $r+1$ are
\emph{$r$-good} (hence ``good'' is short for ``$1$-good'').  A
sequence that is not $r$-good is $r$-bad.  By Dickson's Lemma, every
infinite sequence over $\N^k$ is $r$-good (for any finite $r$), i.e., 
$r$-bad sequences are finite.  Bounding the length of $r$-bad
sequences is helpful in applications where an algorithm does not stop
at the first increasing pair. %

Finding a bound on the length of controlled $r$-bad sequences can
elegantly be reduced to the analysis of plain bad sequences,
another benefit of our ``sum of powers of $\N$'' approach.

Write $\bad_{r,\tau}(t)$ for the maximum length
of $t$-controlled $r$-bad sequences over $\N^\tau$. In this section we
prove the following equality:
\begin{gather}
\label{eq-Lr}
\bad_{r,\tau}(t)= \bad_{r\times\tau}(t)\;.
\end{gather}

For a sequence $\xxx=x_0,x_1,\ldots,x_l$ over some $\N^\tau$, an index
$i=0,1,\ldots,l$ and some $p=1,\ldots,r$, we say that $i$ is
\emph{$p$-good} if there is an increasing subsequence of length $p+1$
\emph{that starts with $x_i$}, i.e., some increasing subsequence
$x_{i_1}\leq x_{i_2}\leq \cdots \leq x_{i_{p+1}}$ with $i_1=i$. The
\emph{goodness} of index $i$ is the largest $p$ such that $i$ is
$p$-good.

For example, consider the following sequence over $\N^2$
{\small\begin{equation*}
\xxx =
\tup{3,1},\,
\tup{5,0},\,
\tup{3,5},\,
\tup{2,4},\,
\tup{2,6},\,
\tup{3,1},\,
\tup{4,5},\,
\tup{2,8}\;.
\end{equation*}}
$\xxx$ can be arranged in layers according to goodness, as in
{\small\begin{equation*}\setlength{\arraycolsep}{0pt}
\begin{array}{lccccccccc}
\text{$2$-good indices: }&
\tup{3,1},&
.&
.&
\tup{2,4},&
.&
.&
.&
.&
\\
\text{$1$-good indices: }&
.&
.&
\tup{3,5},&
.&
\tup{2,6},&
\tup{3,1},&
.&
.&
\\
\text{$0$-good indices: }&
.&
\tup{5,0},&
.&
.&
.&
.&
\tup{4,5},&
\tup{2,8}
\end{array}
\end{equation*}}
This transformation applies to sequences over any wqo. It has two
properties:
\begin{description}
\item[Badness of layers:]\qquad\qquad\quad
Assume that $x_i\leq x_j$ is an increasing pair in $\xxx$. If $x_j$ is
$p$-good then, by definition, $x_i$ is at least $(p+1)$-good. Hence $x_i$ and $x_j$
cannot be in the same goodness layer and every layer is a bad subsequence
of $\xxx$.
\item[Number of layers:]\qquad\qquad\quad If $\xxx$ is $r$-bad, every
  index $i$ is at most $(r-1)$-good and the decomposition requires at
  most $r$ non-empty layers.
\end{description}
If we now see the decomposition as transforming a $t$-controlled $r$-bad
sequence $\xxx$ over $\N^\tau$ into a sequence $\xxx'$ over
$\N^{r\times\tau}$, then $\xxx'$ is $t$-controlled and, as we observed
above, bad. Thus
\begin{equation}
		   \bad_{r,\tau}(t)\leq\bad_{r\times\tau}(t)
\end{equation}
holds in general, proving one half of \eqref{eq-Lr}.
\medskip

For the other half, let $\xxx=x_0,\dots,x_l$ be some
$t$-controlled sequence over $\N^{r\times\tau}$. By collapsing $\N^{r\times
  \tau}$ to $\N^\tau$ in the obvious way, $\xxx$ can be transformed into a
sequence $\yyy$ over $\N^\tau$. The two sequences have same length and same
control. Regarding badness, we can show that $\yyy$ is $r$-bad when $\xxx$
is bad, entailing $l+1\leq \bad_{r,\tau}(t)$ and hence
\begin{equation}
		  \bad_{r\times\tau}(t)\leq \bad_{r,\tau}(t)\;.
\end{equation}
For the proof, assume, by way of contradiction, that $\yyy$ is not $r$-bad,
i.e., is $r$-good. Then it contains an increasing subsequence with $r+1$
elements. By the pigeonhole principle, two of these come from the same
summand in $r\times\tau$, hence $\xxx$ contains an increasing pair and is
good, contradicting our assumption.

%% file: bounding.tex
\section{Upper Bound}
\label{sec-bounding}

\autoref{theo-L-upperbound} gives a bounding function for $\bad$. Define
\begin{gather}
\label{eq-def-Mtau}
	 \ubound_{\tau}(t)\egdef \max_{k\in\tau} \bigl\{1+\ubound_{\tau_{\tup{k,t}}}(t+1)\}\;.
\end{gather}
This inductive definition is well-formed since $\tau_{\tup{k,t}}<_m\tau$
and the multiset ordering is well-founded. Note that $\ubound_\emptyset(t)=0$
since $\max \emptyset=0$. For all $\tau$ and $t$, it holds that
$\bad_\tau(t)\leq \ubound_\tau(t)$.

We first show that the maximum in Eq.~\eqref{eq-def-Mtau} is reached
by always choosing the smallest element of $\tau$
(\autoref{sec-max-ubound}), and then use this characterization to
classify $\ubound$ in the Fast Growing Hierarchy
(\autoref{sec-fgh-ubound}).

\subsection{A Maximizing Strategy for {$\ubound$}}\label{sec-max-ubound}
The next Lemma shows that the maximum of all
$1+\ubound_{\tau_{\tup{k,t}}}(t+1)$ used in Eq.~\eqref{eq-def-Mtau} can always be
obtained by taking $k=\min \tau$. This useful fact leads to a simplified
definition of $\ubound$.
\begin{restatable}{lemma}{lemmin}\label{lem-ubound-min}
Let $k=\min\tau$ and $l\in\tau$. Then $\ubound_{\tau_{\tup{l,t}}}(t+1)\leq
\ubound_{\tau_{\tup{k,t}}}(t+1)$ and, hence,
\begin{align*}
  \ubound_\emptyset(t)&=0\\
  \ubound_\tau(t) &= 1+\ubound_{\tau_{\tup{\min\tau,t}}}(t+1)\quad
  \text{for }\tau\neq\emptyset\;.
\end{align*}
\end{restatable}

%% file: hierarchy-upper.tex
The bounding function $\ubound_\tau$ %
grows very fast with the dimension $k$: $\ubound_{\{3\}}$
is already non-elementary for $f(x)=2x+1$.  \Citet{clote} classified
the upper bounds derived from both his construction and that
of \citeauthor{mcaloon} using the \emph{Fast
Growing Hierarchy} $(\FGH{\alpha})_\alpha$ \citep{fast} for
finite ordinals $\alpha$: for a linear control function, he claimed
his bounding function to reside at the $\FGH{k+6}$ level, and
\citeauthor{mcaloon}'s at the $\FGH{k+1}$ level.  We show in this
section a bounding function in $\FGH{k}$; the results of the next
section entail that this is optimal, since we can find a lower bound
for $\bad_{r\times\{k\}}$ which resides in
$\FGH{k}\backslash\FGH{k-1}$ if $k\geq 2$.

\paragraph{The Fast Growing Hierarchy}
The class $\FGH{k}$ of the Fast Growing Hierarchy is the
closure under substitution and limited recursion of the constant,
sum, projections, and $F_n$ functions for $n\leq k$, where
$F_n$ is defined recursively by\footnote{For simplicity's sake,
we present here a version more customary in the recent literature,
including \citet{mcaloon} and \citet{clote}.
Note however that it introduces a corner case at level 1: in
\citet{fast}, $\FGH{0}\subsetneq\FGH{1}$, the latter being the set of
polynomial functions, generated by $F_1(x)\eqdef(x+1)^2$.} 
\begin{align}
  F_0(x)&\eqdef x+1\label{eq:F0}\\
  F_{n+1}(x)&\eqdef F_n^{x+1}(x)\;,\label{eq:Fn}
\end{align}
where $g^{p}$ denotes the $p$-fold application of a function $g$.
The hierarchy is strict for $k\geq 1$,
i.e.\ $\FGH{k}\subsetneq\FGH{k+1}$,
because $F_{k+1}\notin\FGH{k}$.  For small values of $k$, the
hierarchy characterizes some well-known classes of functions:
\begin{itemize}
  \item $\FGH{0}=\FGH{1}$ contains all the linear functions, like
    $\lambda x.x+3$ or $\lambda x.2x$,
  \item $\FGH{2}$ contains all the elementary functions, like
    $\lambda x.2^{2^{x}}$,
  \item $\FGH{3}$ contains all the tetration functions, like $\lambda
    x.\tetra{2}{x}$, etc.
\end{itemize}
The union $\bigcup_k\FGH{k}$ is the set of primitive-recursive
functions, while $F_\omega$ defined by $F_\omega(x)=F_x(x)$ is an
Ackermann-like non primitive-recursive function; we
call \emph{Ackermannian} such functions that lie in %
$\FGH{\omega}\backslash\bigcup_k\FGH{k}$.  Some further
intuition on the relationship between the functions $f$ in $\FGH{k}$
and $F_k$ for $k\geq 1$ can be gained from the following fact: for
each such $f$,
  there exists a finite $p$ s.t.\ $F_k^p$ majorizes $f$, i.e.\ for
  all $x_1,\ldots,x_n$,
  $f(x_1,\ldots,x_n)<F^p_k(\max(x_1,\ldots,x_n))$
  \citep[Theorem~2.10]{fast}%
  .

Readers might be more accustomed to a variant $(A_k)_k$ of the
$(F_k)_k$ called the \emph{Ackermann Hierarchy} \citep[see
e.g.][]{lowdickson}, and defined by
\begin{align*}
  A_1(x)&\eqdef 2x\\
  A_{k+1}(x)&\eqdef A^x_k(1)\text{ for }k\geq 1\;.
\end{align*}
These versions of the Ackermann functions correspond exactly to
exponentiation of 2 and tetration of 2 for $k=2$ and $k=3$
respectively.  One can check that for all $k,p\geq 1$, there exists
$x_{k,p}\geq 0$ s.t., for all $x\geq x_{k,p}$, $A_k(x)>F^p_{k-1}(x)$,
which contradicts $A_k$ being in $\FGH{k-1}$ by
\citep[Theorem~2.10]{fast}.  %
Conversely, $A_k(x)\leq F_k(x)$ for all $k\geq 1$ and $x\geq 0$, which
shows that $A_k$ belongs to $\FGH{k}\backslash\FGH{k-1}$ for $k\geq
2$.

\paragraph{Main Result}
In this section and in the following one, we focus on classifying in
the Fast Growing Hierarchy the function $\ubound_{r\times\{k\}}$
for some fixed $r$, $k$, and (implicit) $f$.  Here the choice for the
control function $f$ becomes critical, and we prefer therefore the
explicit notation $\uboundd{k}{r}$.

The main result of this section is then
\begin{restatable}{proposition}{propupper}\label{prop:upper}
  Let $k,r\geq 1$ be natural numbers and $\gamma\geq 1$ an ordinal. If
  $f$ is a monotone unary function of $\FGH{\gamma}$ with $f(x)\geq\max(1,x)$
  for all $x$, then $\uboundd{k}{r}$ is in $\FGH{\gamma+k-1}$.
\end{restatable}
One can be more general in the comparison with \citeauthor{mcaloon}'s
proof: his Main Lemma provides an upper bound of the form $G'_{k,f}(d\cdot
f(x)^2)$ for some constant $d$, where in turn his $G'_{k,f}$ function
can be shown to be bounded above by a function in $\FGH{\gamma+k+1}$ when $f$
is in $\FGH{\gamma}$.  The $\FGH{k+1}$ bound for linear functions
reported by \citet{clote} is the result of a specific analysis in
\citeauthor{mcaloon}'s Main Corollary.

%% file: hierarchy-lower.tex
We prove in this section that the upper bound of $\FGH{\gamma+k-1}$ for a
control function $f$ in $\FGH{\gamma}$ is tight if $f$ grows fast
enough.

Let $\lex$ denote the lexicographic ordering over $\N^k$, defined by
\begin{multline*}
\!\!\!\!x=\tup{x[1],\dots,x[k]}<_\text{lex}y=\tup{y[1],\dots,y[k]}\\\equivdef
x[1] < y[1]
\vee\left(x[1]=y[1]\wedge\tup{x[2],\dots,x[k]}<_\text{lex}\tup{y[2],\dots,y[k]}\right)\;.
\end{multline*}
This is a well linear ordering for finite $k$ values, and is coarser
than the natural product ordering.  Let us fix a control function $f$;
we denote by $\llexd{r,k}(t)$ the length of the longest $t$-controlled
$r$-bad sequence for $\lex$ on $\N^k$: this implies that for all $t$
\begin{equation}
  \llexd{r,k}(t) \leq \badd{k}{r}(t)\;.
\end{equation}
We derive in this section an \emph{exact} inductive definition for
$\llex$ in the case $r=1$, and show that it yields large enough lower
bounds for $\bad$ in the case of $f=F_\gamma$.

\begin{figure*}
  \begin{equation*}
    \begin{array}{ccccc}%
      \underbrace{f(t)-1\:f(t)-1\:\cdots\:f(t)-1}
      & \underbrace{f(t)-2,\,f(t)-2,\,\cdots,\,f(t)-2}
      & \cdots
      & \underbrace{0,\,0,\,\cdots,\,0}\\
      \lbound{k}(t)\text{ times}
      & \lbound{k}\!\left(\Lbound{k}(t)\right)\text{ times}
      & 
      & \llexd{k}\!\left(\Lbound{k}^{f(t)-1}(t)\right)\text{ times}
    \end{array}
  \end{equation*}
  \caption{\label{fig-lex-decomposition}The decomposition of bad
    sequences for the lexicographic ordering.}
\end{figure*}
\paragraph{An Inductive Definition for {$\llex$}}
We define our strategy for generating the longest bad controlled
sequence for $\lex$ in $\N^k$ by induction on $k$.  Assume as usual
$f(0)>0$; for $k=1$, the longest $t$-controlled sequence is
\begin{equation*}
  f(t)-1,\; f(t)-2,\;\dots,\; 1,\; 0
\end{equation*}
of length $f(t)$, and we define
\begin{equation}\label{eq:l0}
  \llexd{1}(t)= f(t)\;.
\end{equation}

In dimension $k+1$, we consider the bad sequence where the projection
on the first coordinate is segmented into $f(t)$ constant sections,
such that the projection on the $k$ remaining coordinates of each section
is itself a bad sequence of dimension $k$ following the same strategy.
\begin{example}
  The sequence built by our strategy for $k=2$, $t=3$, and $f(x)=x+1$
  is
  \begin{center}
  \newcolumntype{R}{>{\raggedleft}p{.9em}}
  \centering\setlength{\tabcolsep}{1.5pt}
  {\small\begin{tabular}{c|RRRRRRp{1.1em}RRRRp{1.1em}RRRRp{1.1em}Rr}
    $i$     &0&1&2&3&4&5&$\cdots$&10&11&12&13&$\cdots$&26&27&28&29&$\cdots$&58&59\\
    \hline
    $x_i[1]$
            &3&3&3&3&2&2&$\cdots$& 2& 2& 1& 1&$\cdots$& 1& 1& 0& 0&$\cdots$& 0& 0\\$x_i[2]$
            &3&2&1&0&7&6&$\cdots$& 1& 0&15&14&$\cdots$& 1& 0&31&30&$\cdots$& 1& 0\\
    \hline
    $f(i+t)$&4&5&6&7&8&9&$\cdots$&14&15&16&17&$\cdots$&30&31&32&33&$\cdots$&62&63
  \end{tabular}}\end{center}
  It is composed of four sections, one for each value of the
  first coordinate.  The first section starts at $i=0$ and is of
  length $\lbound{1}(3)=4$, the second starts at $i=4$ and is of
  length $\lbound{1}(7)=8$, the third at $i=12$ with length
  $\lbound{1}(15)=16$, and the last at $i=28$ with length
  $\lbound{1}(31)=32$.  The successive arguments of
  $\lbound{1}$ can be decomposed as sums $t+\lbound{1}(t)$ for the
  previously computed argument $t$:
  \begin{equation*}\begin{array}{r@{\;=\;}c@{\;=\;}l}
     7& 3+4&3+\lbound{1}(3)\\
    15& 7+8&7+\lbound{1}(7)\\
    31&15+16&15+\lbound{1}(15)
  \end{array}\end{equation*}
  simply because at each step the starting index is increased by the
  length of the previous section.\qed
\end{example}
We define accordingly an offset function $o$ by
\begin{equation}
  \Lbound{k}(t)\eqdef t+\llexd{k}(t)\;;
\end{equation}
the strategy results in general in a sequence of the form displayed in
\autoref{fig-lex-decomposition} on the first coordinate.  The obtained
sequence is clearly bad for $\lex$; that it is the longest such
sequence is also rather straightforward by induction: each segment of
our decomposition is maximal by induction hypothesis, and we combine
them using the maximal possible offsets.  Hence  
\begin{equation}\label{eq:lk}
  \lbound{k+1}(t)=\sum_{j=1}^{f(t)}\lbound{k}\!\left(\Lbound{k}^{j-1}(t)\right)\;.
\end{equation}

\begin{remark}
  The lexicographic ordering really yields shorter bad sequences than
  the product ordering, i.e.\ we can have
  $\llexd{k}(t)<\badd{k}{1}(t)$, as can be witnessed by the two
  following sequences for $f(x)=2x$ and $t=1$, which are bad for
  $\lex$ and $\leq$ respectively:
  \begin{equation*}
  {\small\setlength{\arraycolsep}{0.5pt}
  \begin{array}{ccccccccccc}
  \tup{1,1},&\tup{1,0},&\tup{0,5},&\tup{0,4},&\tup{0,3},&\tup{0,2},&\tup{0,1},&\tup{0,0}\\
  \tup{1,1},&\tup{0,3},&\tup{0,2},&\tup{0,1},&\tup{9,0},&\tup{8,0},&\tup{7,0},&\tup{6,0},&\tup{5,0},&\dots,&\tup{0,0}
  \end{array}}
  \end{equation*}
  The first sequence, of length $8=\llexd{2}(1)$, is maximal for $\lex$, and
  shorter than the second, of length $14\leq\badd{2}{1}(1)$.\qed
\end{remark}

\paragraph{Lower Bound for $r$-Bad Sequences}
One can further extend this strategy to give a lower bound on the
length of interleavings of $r$-bad
sequences in $\N^k$, %
by simply concatenating $r$
sequences, each starting with a higher offset.  For instance, for
$r=2$, start with the sequence of length $\lbound{k}(t)$; arrived at
this point, the next sequence reaches length
$\lbound{k}(t+\lbound{k}(t))$.  In general
\begin{equation}
  \lbound{r,k}(t)\geq\sum_{j=1}^r\lbound{k}\!\left(\Lbound{k}^{j-1}(t)\right)\;.
\end{equation}

\begin{restatable}{proposition}{proplower}\label{prop-llex-fgh}
  Let $\gamma\geq 0$ be an ordinal and $k,r\geq 1$ natural numbers.
  Then, for all $t\geq 0$, $\llexd[F_\gamma]{r,k}(t)\geq
  F_{\gamma+k-1}^r(t)$.
\end{restatable}
\begin{remark}
Note that, since
\begin{equation*}
  \llexd[F_\gamma]{r,k}(t)\leq\badd[F_\gamma]{k}{r}(t)\leq\uboundd[F_\gamma]{k}{r}(t)%
  \;,
\end{equation*}
\autoref{prop:upper} %
and
\autoref{prop-llex-fgh} together show that
$\uboundd[F_\gamma]{k}{r}$ %
belongs to
$\FGH{\gamma+k-1}\backslash\FGH{\gamma+k-2}$ if $\gamma\geq 1$ and
$\gamma+k\geq 3$.  One can see that the same holds for
$\llexd[F_\gamma]{k}$, since it is defined by limited primitive
recursion.\qed
\end{remark}
\begin{remark}
  In the case of the successor control function $f=F_0$, the $F_{k-1}$
  lower bound provided by \autoref{prop-llex-fgh} does not
  match the $\FGH{k}$ upper bound of \autoref{prop:upper}
  (indeed the statement of the latter does not allow $\gamma=0$ and
  forces $\gamma=1$).  Tightness holds nevertheless, since
  \citet{lowdickson} proved in his Theorem~2.6 an $A_k$ lower bound
  for this particular case of $f=F_0$.\qed
\end{remark}

\paragraph{Concrete Example}
It is easy to derive a concrete program illustrating the intuition behind
\autoref{prop-llex-fgh}:
\begin{example}\label{ex:lex}
  Consider the following program with control
  $\lambda x.2^x+1$ in $\FGH{2}$ for $t=\lceil\log_2\max_{1\leq j\leq
    k} a_j\rceil$%
  :\\[.4em]
  {\small\begin{tabular}{l}
      \textsc{lex} $(a_1,\dots,a_k)$\\
      $c\longleftarrow 1$\\
      \textbf{while} {$\bigwedge_{1\leq j\leq k} a_j>0$}\\
      \qquad $\tup{a_1,a_2,\ldots,a_{k-1},a_k,c}\longleftarrow\tup{a_1-1,2c,\ldots,2c,2c,2c}$\\
      \quad\textbf{or}\\
      \qquad $\tup{a_1,a_2,\ldots,a_{k-1},a_k,c}\longleftarrow\tup{a_1,a_2-1,\ldots,2c,2c,2c}$\\
      \quad\textbf{or}\\
      \quad~$\vdots$\\
      \quad\textbf{or}\\
      \qquad $\tup{a_1,a_2,\ldots,a_{k-1},a_k,c}\longleftarrow\tup{a_1,a_2,\ldots,a_{k-1},a_k-1,2c}$\\
      \textbf{end}
\end{tabular}}\\[-.3em]

An analysis similar to that of $\llexd{k}$ shows that, for
$k\geq 2$ and $m=\min_{1\leq j\leq k} a_j>0$, \textsc{lex} might run
through its \textbf{while} loop more than $A_{k+1}(m)$ times, which is
a function in $\FGH{k+1}\backslash\FGH{k}$.  It matches the $\FGH{k+1}$
upper bound provided by \autoref{prop:upper} for this program, since
the projection of any sequence of program configurations
$\tup{a_1,\ldots,a_k,c}$ on the $k$ first components is bad ($c$
increases continuously and thus does not contribute to the sequence
being bad).\qed
\end{example}

%% file: applications.tex
Results on the length of bad sequences %
are rarely used in the verification
literature.
We claim that \autoref{prop:upper} is very easy to
use when one seeks complexity upper bounds, at least if one is content
with the somewhat coarse bounds provided by the Fast Growing
Hierarchy.%
One might want to modify the choices of parametrization we
made out of technical convenience: for instance
\begin{itemize}
\item controlling the sum of the vector components instead of their
  infinity norm, i.e.\ asking that $\sum_j x_i[j]<f(i+t)$: since
  $|x_i|_\infty\leq\sum_j x_i[j]$, \autoref{prop:upper} also works for
  this definition of control,
\item controlling the bitsize of the successive vectors in a bad
  sequence similarly only induces a jump in the classification of $f$
  from $\FGH{1}$ to $\FGH{2}$ and leaves the other cases unchanged,
\item using an ``internal'' view of the control, constraining how much
  the vector components can grow in the course of a single step of the
  algorithm, i.e.\ such that $|x_i|_\infty < f^i(t)$, leads to upper
  bounds one level higher in the Fast Growing Hierarchy, since
  $\lambda i.f^{i+1}(t)$ controls the sequence in our sense and
  belongs to $\FGH{\gamma+1}\!$ whenever $f$ belongs to $\FGH{\gamma}$.
\end{itemize}

\subsection{Disjunctive Termination Arguments}\label{sub:rank}
Program termination proofs essentially establish that the program's
transition relation $R$ is well-founded.  The classical,
``monolithic'' way of proving well-foundedness is to exhibit a
\emph{ranking} function $\rho$ from the set of program configurations
$x_0,x_1,\ldots$ into a well-order such that
$R\subseteq\{(x_i,x_j)\mid \rho(x_i)\not\leq\rho(x_j)\}$, like
$\lambda a_1\cdots a_kc.(\sum_{1\leq j\leq k}\omega^{k-j+1}\cdot a_j)$,
mapping $\N^{k+1}$ to $\omega^k$ for \autoref{ex:lex}.
That same ranking function could also be seen as mapping to $(\N^k,\lex)$, a linear extension of
the product ordering.  Our techniques easily apply to such termination
proofs based on lexicographic orderings: one only needs to identify a
control function. This is usually obtained by combining the computational
complexities of the  program operations and of the ranking
function.

A different termination argument was proposed by
\citet{PodelskiR-lics04} \citep[see also][]{blass08,cook11}: in order
to prove $R$ to be well-founded, they rather exhibit a finite set of
well-founded relations $T_1,\dots,T_k$ and prove that $R^+\subseteq
T_1\cup\cdots\cup T_k$.  In practice, each of the $T_j$, $1\leq j\leq
k$, is proved well-founded through a ranking function $\rho_j$, but
these functions might be considerably simpler than a monolithic
ranking function.  In the case of \autoref{ex:lex}, choosing
$T_j=\{(\tup{a_1,\ldots,a_j,\ldots,a_k,c},\tup{a'_1,\ldots,a'_j,\ldots,a'_k,c'})\mid
a_j> 0\wedge a_j'<a_j\}$, yields such a \emph{disjunctive termination
argument}.

Although \citeauthor{PodelskiR-lics04} resort to Ramsey's Theorem in
their termination proof, we can easily derive an alternative proof
from Dickson's Lemma, which allows us to apply our results: if
each of the $T_j$ is proven well-founded thanks to a mapping $\rho_j$
into some wqo $(X_j,\leq_j)$, then with a sequence
$x_0,x_1,\ldots$
of program configurations one can associate the sequence of tuples
  $\tup{\rho_1(x_0),\ldots,\rho_k(x_0)},\tup{\rho_1(x_1),\ldots,\rho_k(x_1)},\ldots$
in $X_1\times\cdots\times X_k$, the latter being a wqo for the product
ordering by Dickson's Lemma.  Since for any indices $i_1<i_2$,
$(x_{i_1},x_{i_2})\in R^+$ is in some $T_j$ for some $1\leq j\leq k$,
we have $\rho_j(x_{i_1})\not\leq_j\rho_j(x_{i_2})$ by definition of a
ranking function.  Therefore the sequence of tuples is bad for the
product ordering and thus finite, and the program terminates.

If the range of the ranking functions is $\N$, one merely needs to
provide a control on the ranks $\rho_j(x_i)$, i.e.\ on the composition
of $R^i$ with $\rho_j$, in order to apply \autoref{prop:upper}.
For instance, for all programs consisting of a loop with variables ranging
over $\mathbb{Z}$ and updates of linear complexity (like \textsc{choice}
or \textsc{lex}),
\citet{BradleyMS-concur05} synthesize linear ranking functions into
$\N$:
\begin{question}\label{q-rank}
  What is the complexity of loop programs with linear operations
  proved terminating thanks to a \mbox{$k$-ary} disjunctive termination
  argument that uses linear ranking functions into $\N$?
\end{question}\noindent
The control on the ranks in such programs is at most exponential (due
to the iteration of the loop) in $\FGH{2}$. With \autoref{prop:upper} one obtains an upper
bound in $\FGH{k+1}$ on the maximal number of loop iterations (i.e., the
running time of the program), where
$k$ is the number of transition invariants $T_1,\ldots,T_k$ used in
the termination proof---in fact we could replace ``linear'' by
``polynomial'' in \autoref{q-rank} and still provide the same answer.
\autoref{ex:lex} shows this upper bound to be tight.  Unsurprisingly,
our bounds directly relate the complexity of programs with the number
of disjunctive termination arguments required to prove their
termination.

\subsection{Reachability for Incrementing Counter Automata}\label{sub:xpath}
\emph{Incrementing Counter Automata}, or ICA's, are Minsky counter machines
with a modified operational semantics \cite[see][]{Demri-jancl06,demri09}.
ICA's have proved useful for deciding logics on data words and data trees, like
\textsc{XPath} fragments~\citep{figueira09}. The fundamental
result in this area is that, for ICA's, the set of reachable
configurations is a computable set~\citep{mayr2003,phs-rp2010}.

Here we only introduce a few definitions and notations that are
essential to our development (and refer to~\citep{mayr2003,phs-rp2010}
for more details). The configuration of a $k$-counter machine
$M=(Q,\Delta)$ is some tuple $v=\tup{q,a_1,\ldots,a_k}$ where $q$ is a
control-state from the finite set $Q$, and $a_1,\ldots,a_k\in \N$ are
the current values of the $k$ counters. Hence $\Conf_M\egdef
Q\times \N^k$. The transitions between the configurations of $M$ are
obtained from its rules (in $\Delta$). Now, whenever $M$ seen as a
Minsky machine has a transition
$\tup{q,a_1,\ldots,a_k}\Mstep\tup{p,b_1,\ldots,b_k}$, the same $M$
seen as an ICA has all transitions $\tup{q,a_1,\ldots,a_k}\Istep
\tup{p,b'_1,\ldots,b'_k}$ for $b'_1\geq b_1\wedge\cdots \wedge b'_k\geq
b_k$: Informally, an ICA behaves as its underlying Minsky machine,
except that counters may increment spuriously after each step. The
consequence is that, if we order $\Conf_M$ with the standard partial
ordering (by seeing $\Conf_M$ as the wqo $\sum_{q\in Q}\N^k$), then
the reachability set of an ICA is upward-closed.
\medskip

We now describe the forward-saturation algorithm that computes the
reachability set from an initial configuration $v_0$.

Let $X_0,X_1,X_2,...$ and $Y_0,Y_1,Y_2,...$ be the sequences of subsets of
$\Conf_M$ defined by
\begin{xalignat*}{2}
X_0&\egdef\{v_0\},
&
X_{i+1}&\egdef \Post(X_i),
\\
Y_0&\egdef X_0,
&
Y_{i+1}&\egdef Y_i\cup X_{i+1},
\end{xalignat*}
where
$
\Post(X)\egdef\{v'\in \Conf_M~|~\exists v\in X:\: v\Istep v'\}.
$
The reachability set is $\Reach(M,v_0)\egdef \bigcup_{i=1,2,\ldots}X_i$,
i.e., $\lim_{i\rightarrow\omega} Y_i$. However, since every $X_{i+1}$ is
upward-closed, the sequence $(Y_i)_{i\in\N}$ stabilizes after finitely many
steps, i.e., there is some $l$ such that
\mbox{$Y_l=Y_{l+1}=\cdots=\Reach(M,v_0)$}, as we prove below. %
This method is effective once we
represent (infinite) upward-closed sets by their finitely many minimal
elements: it is
easy to compute the minimal elements of $X_{i+1}$ from the
minimal elements of $X_i$, hence one can build the sequence $Y_0, Y_1,
\ldots$ (again represented by minimal elements) until
stabilization is detected.

\begin{question}
What is the computational complexity of the above forward-saturation
algorithm for ICA's?
\end{question}\noindent
For this question, we start with the length of the sequence
$Y_0\varsubsetneq Y_1 \varsubsetneq Y_2 \varsubsetneq \cdots
\varsubsetneq Y_l= Y_{l+1}$. For each $i=1,\ldots,l$, let $v_i$ be a
minimal element in $Y_i\setminus Y_{i-1}$ (a \emph{nonempty} subset of
$\Conf_M$). Note that $v_i\in X_i$, an upward-closed set, so that
$Y_i$ contains all configurations above $v_i$. Hence $v_j\not\geq v_i$
for $j>i$ (since $v_j\not\in Y_i$) and the sequence
$\vvv=v_1,v_2,\ldots$ is bad---this also proves the
termination of the $(Y_i)_i$ sequence.

We now need to know how $\vvv$ is controlled.
Consider a minimal element $v$ of $Y_i$.  Then $\leni{v}\leq
i+\leni{v_0}$, which means that $\vvv$ is $\leni{v_0}$-controlled for
$f=F_0$ the successor function.
Here $f$ is independent of the ICA $M$ at hand!
Using \autoref{prop:upper} we conclude that, for fixed $k$, $l$ is bounded
by a function in $\FGH{k}$ with $\leni{v_0}$ as argument. Now, computing
$X_{i+1}$ and $Y_{i+1}$ (assuming representation by minimal elements) can
be done in time linear in $\len{X_i}$ and $\len{Y_i}$ (and $\len{M}$ and $\leni{v_0}$),
so that the running time of the algorithm is in $O(\len{M}\cdot l)$, i.e.,
also in $\FGH{k}$ \citep[see][for $F_{k-2}$ lower bounds for
the reachability problem in $k$-dimensional ICA's]{phs-mfcs2010}.

Here the main parameter in the complexity is the number $k$ of
counters, not the size of $Q$ or the number of rules in $M$. For fixed $k$
the complexity is primitive-recursive, and it is
Ackermannian when $k$ is part of the input---which is the case in
the encoding of logical formul\ae\ of \citet{demri09}.

\subsection{Coverings for Vector Addition Systems}\label{sub:km}

Vector addition systems (VAS's) are systems where $k$ counters evolve by
non-deterministically applying $k$-dimensional translations from a fixed
set. They can be seen as an abstract presentation of Petri nets, and are
thus widely used to model concurrent systems, reactive systems with
resources, etc.

Formally, a $k$-dimensional VAS is some $S=(\Delta,v_0)$ where
$v_0\in \N^k$ is an \emph{initial configuration} and $\Delta\subseteq
\Z^k$ is a finite set of \emph{translations}. Unlike translations,
configurations only contain non-negative values. A VAS $S$ has a step
$v\step{\delta} v'$ whenever $\delta\in\Delta$ and $v+\delta\in\N^k$:
we then have $v'=v+\delta$. Hence the negative values in $\delta$ are
used to decrement the corresponding counters \emph{on the condition
  that they do not become negative}, and the positive values are used
to increment the other counters. A configuration $v$ is reachable,
denoted $v\in \Reach(S)$, if there exists a sequence
$
v_0\step{\delta_1}v_1\step{\delta_2}v_2 \cdots \step{\delta_n}v_n=v \:.
$
That reachability is decidable for VAS's is a major result of computer
science but we are concerned here with computing a \emph{covering} of the
reachability set.

In order to define what is a ``covering'', we consider the completion
$\N_\omega\egdef\N\cup\{\omega\}$ of $\N$ and equip it with the
obvious ordering. Tuples $w\in\N_\omega^k$,
called \emph{$\omega$-markings}, are ordered with the product
ordering.  While $\omega$-markings are not proper configurations, it
is convenient to extend the notion of steps and write
$w\step{\delta}w'$ when $w'=w+\delta$ (assuming $n+\omega=\omega$ for
all $n$).

Let $C\subseteq \N_\omega^k$ be a set of $\omega$-markings. We say
that $C$ is a \emph{covering for $S$} if %
for any $v\in\Reach(S)$, $C$ contains some $w$ with $v\leq w$, while any
$w\in C$ is in the adherence of the reachability set, i.e.,
$w=\lim_{i=1,2,\ldots} v_i$ for some markings
$v_1,v_2,\ldots$ in $\Reach(S)$.
Hence a covering is a rather precise approximation of the reachability
set (precisely, the adherence of its downward-closure). A fundamental
result is that \emph{finite} coverings always exist and are
computable.  This entails several decidability results, e.g.\ whether
a counter value remains bounded throughout all the possible runs.%
\medskip

A particular covering of $S$ can be obtained from the KM tree,\footnote{
  The computation of the KM tree has other uses, e.g., with the finite
  containment problem~\cite{frt}. Results from \citet{frt} %
  show Ackermannian lower bounds, and provided the initial motivation for
  the work of \citet{mcaloon} and \citet{clote}.
} introduced by \citet{kmtree}.
Formally, this tree has nodes labeled with $\omega$-markings and edges
labeled with translations. The root $s_0$ is labeled with $v_0$ and the tree is
grown in the following way:
Assume a node $s$ of the tree is labeled with some $w$ and let
  $(v_0=)w_0, w_1,..., w_n=w$ be the labels on the path from the root to
  $s$. For any translation $\delta\in\Delta$ such that there is a step
  $w\step{\delta}w'$, we consider whether to grow the tree by adding a
  child node $s'$ to $s$ with a $\delta$-labeled edge from $s$ to $s'$.
\begin{enumerate}
\item\label{itm2}
If $w'\leq w_i$ for one of the $w_i$'s on the path from $s_0$ to $s$, we
do not add $s'$ (the branch ends).
\item\label{itm3} Otherwise, if $w'>w_i$ for some $i=0,\ldots,n$, we build $w''$ from $w'$
  by setting, for all $j=1,\ldots,k$, $w''[j]\egdef\omega$ whenever
  $w'[j]>w_i[j]$, otherwise $w''[j]$ is just $w'[j]$. Formally, $w''$ can
  be thought as ``$w_i+\omega\times(w'-w_i)$''. We add $s'$, the edge
  from $s$ to $s'$, and we label $s'$ with $w''$.
\item\label{itm4} Otherwise, $w'$ is not comparable with any $w_i$: we simply add the edge and label $s'$ with $w'$.
\end{enumerate}
\begin{theorem}[\citep{kmtree}]
The above algorithm terminates and the set of labels in the KM tree is
a covering for $S$.
\end{theorem}

\begin{question}
What is the %
complexity of the KM algorithm? What is the
size of the KM tree? And the size of $C$?
\end{question}
\noindent
Answering the above question requires understanding why the KM
algorithm terminates. First observe that the KM tree is finitely
branching (a node has at most $\len{\Delta}$ children), thus the tree
can only be infinite by having an infinite branch (K\H{o}nig's
Lemma). Assume, for the sake of contradiction, that there is an
infinite branch labeled by some $w_0,w_1,\ldots$ The sequence may be a
good sequence, but any increasing pair $w_{i_1}\leq w_{i_2}$ requires
$w_{i_2}$ to be inserted at step~\ref{itm3} of the KM algorithm.
Hence $w_{i_2}$ has more $\omega$'s than $w_{i_1}$. Finally, since an
$\omega$-marking has at most $k$ $\omega$'s, the sequence is
$(k+1)$-bad and cannot be infinite since $\N_\omega^k$ is a wqo.

Now, how is the sequence controlled? If we say that the $\omega$'s do not
count in the size of an $\omega$-marking, a branch $w_0,w_1,\ldots$ of the
KM tree has $\leni{w_{i+1}}\leq
\leni{w_i}+\leni{\Delta}\leq\leni{v_0}+i\cdot\leni{\Delta}$.  Hence the
sequence is $\leni{v_0}$-controlled for $f(x)=x\cdot\leni{\Delta}+1$, a
control at level $\FGH{1}$ for fixed $\Delta$.  More coarsely, the sequence
is $\len{S}$-controlled for a \emph{fixed} $f(x)=x^2$, this time at level
$\FGH{2}$.  By \autoref{prop:upper} and Eq.~\eqref{eq-Lr}, we deduce that the
length of any branch is less than
$l_{\max}=\bad_{(k+1)\times\{k\}}(\len{S})$. The size of the KM tree, and
of the resulting
$C$, is bounded by $\len{\Delta}^{l_{\max}}$. Finally, the
time complexity of the KM algorithm on $k$-dimensional VAS's\linebreak is in $\FGH{k+1}$: the complexity is primitive-recursive for fixed
dimensions, but Ackermannian when $k$ is part of the input.

The above result on the size of KM trees can be compared with the
tight bounds that \citeauthor{howell86} show for
VAS's~\cite[Theorem~2.8]{howell86}.  Their $\FGH{k-1}$ bound is two
levels better than ours.  It only applies to KM trees and is obtained
via a rather complex analysis of the behaviour of VAS's, not a generic
analysis of Dickson's Lemma. In particular it does not apply to VAS
extensions, while our complexity analysis carries over to many classes
of well-structured counter systems, like the strongly increasing
affine nets of~\cite{finkel-wsn}, for which both the KM tree algorithm
and a $\FGH{2}$ control keep
applying%
, and thus so does the $\FGH{k+1}$ bound.%

%% file: related.tex
\paragraph{Bounds for $\N^k$}

We are not the first ones to study the length of controlled bad
sequences.  Regarding Dickson's Lemma, both \citet{mcaloon} and
\citet{clote} employ \emph{large intervals} in a sequence and their
associated Ramsey theory, showing that large enough intervals would
result in good sequences.  Unlike our elementary argument based on
disjoint sums, we feel that the combinatorial aspects of
\citeauthor{mcaloon}'s approach are rather complex, whereas the
arguments of \citeauthor{clote} %
rely on a long
analysis performed by \citet{ramseyfunc} and is not parametrized by
the control function $f$.  Furthermore, as already mentioned on
several occasions, both proofs result in coarser upper bounds.
\Citet[Theorem~6.2]{lowdickson} also shows that bad sequences over $\N^k$ are
primitive-recursive but the proof is given for the
specific case of the successor function as control, and does not
distinguish the dimension $k$ as a parameter.  One could also see the
results of \citet{howell86} or \citet{hofbauer92} as implicitly
providing bounds on the bad sequences that can be generated resp.\ by
VAS's and certain terminating rewrite systems; using these bounds for
different problems can be cumbersome, since not only the control
complexity is fixed, but it also needs to be expressed in the formal
system at hand.

\paragraph{Beyond $\N^k$}

Bounds on bad sequences for other wqo's have also been considered;
notably \citet{cichon98} provide bounds for finite sequences with the
embedding order (Higman's Lemma).  Their bounds use a rather complex
ordinal-indexed hierarchy.  If we only consider tuples of natural
numbers, their decomposition also reduces inductively from $\N^k$ to
$\N^{k-1}$, but it uses the ``badness'' parameter ($r$, see~\autoref{sec-long-r-bad}) as a useful tool, as
witnessed by their exact analysis of $\bad_{r,1,f}$.  For arbitrary $k\in\Nat$,
\citeauthor{cichon98} have an elegant decomposition, somewhat similar
to the large interval approach, that bounds
$\bad_{r,k,f}$ by some $\bad_{r',k-1,f'}$ for some $r'$ and $f'$
obtained from $r$, $f$ and $k$.  However, $r'$ and $f'$, $r''$ and
$f''$, \dots, quickly grow very complex, and how to classify the
resulting bounds in the Fast Growing Hierarchy is not very clear to
us.  By contrast, our approach lets us keep the same fixed control
function $f$ at all steps in our decomposition, and it can handle Higman's
Lemma as demonstrated in~\cite{SS-arxiv11}.  %

\Citeauthor{weiermann94} proves another bound for Higman's Lemma
\citep[Corollary~6.3]{weiermann94}, but his main focus is actually to
obtain bounds for Kruskal's Theorem~\citep[Corollary~6.4]{weiermann94},
i.e.\ for finite trees with the embedding ordering.  The bounds
are, as expected, very high, and only consider polynomial
ranking functions.

\paragraph{Further Pointers}
The question of extracting complexity upper bounds from the use of
Dickson's Lemma can be seen as an instance of a more general concern
stated by Kreisel: ``What more than its truth do we know if we have a
proof of a theorem in a given formal system?''  Our work fits in the
field of implicit computational complexity in a broad sense, which
employs techniques from linear logic, lambda calculus and typing,
invariant synthesis, term rewriting, etc. that entail complexity
properties.  In most cases however, the scope of these techniques is
very different, as the complexity classes under study are quite low
with e.g. \textsc{PTime} being the main object of
focus \citep[e.g.][etc.]{leivant02,gulwani09,hofmann10}.
By contrast, our technique is of
limited interest for such low complexities, as the Fast Growing
Hierarchy only provides very coarse bounds.  But it is well suited for
the very large complexities of many algorithmic issues, for
well-structured transition systems~\citep{FinSch-WSTS} working on
tuples of naturals, Petri nets equivalences~\citep{frt,jancar},
Datalog with constraints~\citep{revesz93}, Gr\"obner's
bases~\citep{gallo94}, relevance logics~\citep{urquhart99}, LTL with
Presburger constraints~\citep{Demri-jancl06}, data
logics~\citep{demri09,figueira09}, etc.

A related concept is the \emph{order type} of a well partial
order~\citep{dejongh77}, which roughly corresponds to the maximal
transfinite length of an uncontrolled bad sequence.  Although order
types do not translate into bounds on controlled
sequences,\footnote{For instance, $\omega^k$ is the order type of both
$(\N^k,\leq)$ and $(M(\Sigma_k),\subseteq)$, where $M(\Sigma_k)$ is
the set of multisets over a finite set $\Sigma_k$ with $k$ elements,
but one needs to be careful on how a control on one structure
translates into a control for the other.} they are sometimes good
indicators, a rule of thumb being that an upper bound in
$\FGH{\alpha}$ is often associated with an order type of
$\omega^\alpha$, which actually holds in our case.  Such questions
have been mostly investigated for the complexity of term rewriting
systems~\citep[see][and the references therein]{lepper04}, where for
instance the maximal derivation length of a term rewriting system \enlargethispage{.1em}
compatible with multiset termination ordering (of order $\omega^k$ for
some finite $k$) was shown primitive-recursive by \citet{hofbauer92}
(however no precise bounds in terms of $k$ were given).

%% file: conclusion.tex
\section{Conclusion}
In spite of the prevalent use of Dickson's Lemma in various areas of
computer science, the upper bounds it offers are seldom capitalized
on.  Beyond the optimality of our bounds in terms of the Fast Growing
Hierarchy, our first and foremost hope is for our results to improve
this situation, and reckon for this on
\begin{itemize}
\item an arguably simpler main proof argument, that relies on a
  simple decomposition using disjoint sums,
\item a fully worked out classification for our upper bounds---a
  somewhat tedious task---, which is reusable because we leave the
  control function as an explicit parameter,
\item three template applications where our upper bounds on bad
  sequences translate into algorithmic upper bounds. These are varied
  enough not to be a mere repetition of the exact same argument,
  and provide good illustrations of how to employ our results.
\end{itemize}

\section*{Acknowledgment}
\addcontentsline{toc}{section}{Acknowledgment}
The authors gratefully acknowledge the contribution of an anonymous
reviewer, who pointed out the application to transition invariants
given in \autoref{sub:rank}.

%% file: app-decomposition.tex
\subsection{Proof of Theorem~\ref{theo-L-upperbound}}\label{ax:L-upperbound}
\Lupperbound*

We start with some necessary notation and basic facts: For two
quasiorderings $(A_1,\leq_1)$ and $(A_2,\leq_2)$, a mapping
$h:A_1\rightarrow A_2$ is a \emph{reflection} when
\begin{equation*}
  \forall a,b\in A_1: h(a)\leq_2 h(b) \text{ implies } a\leq_1 b \, .
\end{equation*}
We further say that it is a \emph{strong reflection} when $\leni{h(x)}\leq
\leni{x}$ for all $x$. (NB: we only consider reflections between
quasiorderings that are subsets of some $\N^\tau$, hence the notion of size
is well-defined.) When $h$ is a strong reflection, we write
$h:A_1\hookrightarrow A_2$ (or just $A_1\hookrightarrow A_2$ when $h$ is
left implicit) and say that $A_2$ \emph{strongly reflects} $A_1$.

Strong reflections preserve controlled bad sequences: assume
$h:A_1\hookrightarrow A_2$ and that $x_0,x_1,\ldots,x_l$ is a
$t$-controlled bad sequence over $A_1$. Then $h(x_0),h(x_1),\ldots,h(x_l)$
is a $t$-controlled bad sequence over $A_2$.

This notion is compatible with the composition of orderings:
\begin{fact}
Let $A,A_1,A_2$ be  quasiorderings:
$A_1\hookrightarrow A_2$ { implies }
$A+A_1 \hookrightarrow A+A_2${ and }
$A\times A_1\hookrightarrow A\times A_2$.
\end{fact}

For $a\in A$, we let $A / a\egdef\{x\in A~|~a\not\leq x\}$ denote the
subset of elements that are not above $a$. Note that $(A/b)\subseteq(A/a)$
when $a\leq b$.

When $(A,\leq)$ is a wqo, $(A/a,\leq)$ is clearly a wqo
too, called a \emph{residual wqo}.
The point is that if $\xxx=x_0,x_1,\ldots$ is a bad sequence over
some $A$, the suffix sequence $\yyy=x_1,\ldots$ is a bad sequence over
$A/x_0$. In the following, we extend our notations and write $\bad_{A}(t)$
for the maximal length of a $t$-controlled bad sequence over $A$ when $A$
is a subset of some $\N^\tau$.

Here too, the notion of residuals is compatible with the composition of orderings:
if $a$ is in $A_j$, we have for a disjoint sum $\sum_{i\in I} A_i$ with
$j\in I$
\begin{equation}
\label{eq-hook-sum}
(\sum_{i\in I} A_i) / a =
(A_j/a)+\sum_{i\in I\backslash\{j\}} A_i\;.
\end{equation}
More crucially, the region-based decomposition of \autoref{sec-dickson}
relies on a reflection for products
\begin{equation}
\bigl((A\times B) / \tup{a,b}\bigr)
\hookrightarrow
\bigl((A / a)\times B \,+\, A \times (B / b) \bigr) \, .
\end{equation}
An immediate corollary is
\begin{equation}
\label{eq-hook-pow}
(A^k / \tup{a,\ldots,a}) \hookrightarrow k\times (A / a)\times A^{k-1} \, .
\end{equation}

\begin{lemma}
\label{lem-Nk-hook}
Assume $x\in\N^k$ with $k>0$ and $\leni{x}\leq f(t)-1$:
\begin{equation*}
\label{eq-refl-k}
	   \N^k	 / x \hookrightarrow k\times(f(t)-1)\times \N^{k-1} 
\;\;(\text{i.e., } N_k(t)\times\N^{k-1})  \, .
\end{equation*}
\end{lemma}\noindent
Indeed, when $k=1$, $\N / x = \{0, 1, \ldots, x-1\}$, which is
isomorphic to $x\times\N^0$, in turn strongly reflected by
$(f(t)-1)\times\N^0$, while for $k>1$ we reduce to the 1-dimensional
case using Eq.~\eqref{eq-hook-pow}.

By definition of $\tau_{\tup{k,t}}$ (see Eq.~\ref{eq-def-taukt}), combining \autoref{lem-Nk-hook} and
Eq.~\eqref{eq-hook-sum} directly yields
\begin{lemma}
\label{lem-Ntau-hook}
Assume	$k\in \tau$ and $x\in\N^k$ with $\leni{x}\leq f(t)-1$:
\begin{equation*}
\label{eq-refl-tau}
	   \N^\tau / x \hookrightarrow \N^{\tau_{\tup{k,t}}} \, .
\end{equation*}
\end{lemma}
Since strong reflections preserve controlled bad sequences,
we deduce
\begin{equation}
\label{eq-hook-to-bad}
{A_1}\hookrightarrow {A_2} \text{ implies } \bad_{A_1}(t)\leq \bad_{A_2}(t) \, 
\end{equation}
where, for $i=1,2$, $A_i$ is some $\N^{\tau_i}$, or one of its residuals.

We are now sufficiently equipped.
\begin{proof}[Proof (of \autoref{theo-L-upperbound})]
The proof is by induction over $\tau$, the base case $\tau=\emptyset$
holding trivially in view of $\bad_{\emptyset}(t)=0$.  For the inductive
case, assume $\tau\not=\emptyset$ and let $\xxx=x_0,x_1,\ldots,x_l$ be a
$t$-controlled bad sequence over $\N^\tau$ with maximal length, so that
$\bad_{\tau}(t)=l+1$. Write $\yyy=x_1,\ldots,x_l$ for the suffix sequence:
$\yyy$ is a $(t+1)$-controlled bad sequence over $\N^{\tau} / x_0$. Since
$x_0$ belongs to $\N^k$ for some $k\in\tau$, we deduce $l\leq
\bad_{\tau_{\tup{k,t}}}(t+1)$ by combining \autoref{lem-Ntau-hook} and
Eq.~\eqref{eq-hook-to-bad} and using the induction hypothesis.  Which
concludes our proof.
\end{proof}

%% file: hierarchy-computations.tex
\subsection{Proof of Lemma~\ref{lem-ubound-min}}\label{ax:min-strategy}
Let us first introduce a third, less standard, so-called
``\emph{dominance}'' ordering on multisets, given by
\begin{multline}
\{a_1,\ldots,a_n\}\sqsubseteq
\{b_1,\ldots,b_m\}
\equivdef
n\leq m \wedge a_1\leq b_1 \wedge \ldots \wedge a_n\leq b_n
\end{multline}
where it is assumed that elements are denoted in decreasing order, i.e.,
$a_1\geq a_2\geq \ldots\geq a_n$ and $b_1\geq \ldots \geq b_m$. In other
words, $\tau\sqsubseteq\tau'$ when every element in $\tau$ is dominated by
its own sibling element in $\tau'$ (additionally $\tau'$ may have extra
elements). For dominance, reflexivity and transitivity are clear. We let
the reader check that the dominance ordering sits between the inclusion
ordering and the multiset ordering.

In order to exploit Eq.~\eqref{eq-def-Mtau}, we need some basic
properties of the operation that transforms $\tau$ into $\tau_{\tup{k,t}}$.
\begin{lemma}[About $\tau_{\tup{k,t}}$]
\label{lem-tau-k-t}~\hfill
\begin{enumerate}
\item\label{lem-tau-k-t-1} $\tau_{\tup{k,t}}<_m\tau$.
\item\label{lem-tau-k-t-2} If $\tau\subseteq\tau'$ then
  $\tau_{\tup{k,t}}\subseteq\tau'_{\tup{k,t}}$.
\item\label{lem-tau-k-t-3} If $\{k,l\}\subseteq\tau$ then
  $\tau_{\tup{l,t}\tup{k,t'}}=\tau_{\tup{k,t'}\tup{l,t}}$.
\item\label{lem-tau-k-t-4} If $\{k,l\}\subseteq\tau$ with furthermore
  $k\leq l$ and $t\leq t'$, then
  $\tau_{\tup{l,t}\tup{k,t'}}\sqsubseteq\tau_{\tup{k,t}\tup{l,t'}}$.
\item\label{lem-tau-k-t-5} If $\tau\sqsubseteq\tau'$ and $k\in\tau$
  then there exists $l\in\tau'$ such that $k\leq l$ and
  $\tau_{\tup{k,t}}\sqsubseteq\tau'_{\tup{l,t}}$.
\end{enumerate}
\end{lemma}
\begin{proof}[Proof Sketch]
For~\textit{\ref{lem-tau-k-t-3}}, we note that $\tau_{\tup{l,t}\tup{k,t'}}$ and
$\tau_{\tup{k,t'}\tup{l,t}}$ are obtained from $\tau$ by performing
\emph{exactly the same removals and additions of elements}. These are
perhaps performed in different orders, but this does not change the end
result.
\\

For~\textit{\ref{lem-tau-k-t-4}}, we note that $\tau_{\tup{l,t}\tup{k,t'}}$ is some
$\tau-\{k,l\}+\tau_1$ for
\[
	    \tau_1 = N_{l}(t)\times\{l-1\}+N_{k}(t')\times\{k-1\}
\]
while $\tau_{\tup{k,t}\tup{l,t'}}$ is $\tau-\{k,l\}+\tau_2$ for
\[
	   \tau_2 = N_{l}(t')\times\{l-1\}+N_{k}(t)\times\{k-1\}\;.
\]
From $l\geq k$ and $f(t)\leq f(t')$ we deduce
\[
		  N_{l}(t)+N_{k}(t')\leq N_{l}(t')+N_{k}(t)\;.
\]
Hence $\tau_1$ has less elements than $\tau_2$. Furthermore, $\tau_1$ has
less of the larger ``$l-1$'' elements since $N_{l}(t)\leq N_{l}(t')$. Thus
$\tau_1\sqsubseteq\tau_2$, entailing
$\tau-\{k,l\}+\tau_1\sqsubseteq\tau-\{k,l\}+\tau_2$.
\\

For~\textit{\ref{lem-tau-k-t-5}}, we use the $l=b_i$ that corresponds to
$k=a_i$ in the definition of dominance ordering. This ensures both
$k\leq l$ (hence $N_{k}(t)\leq N_{l}(t)$ and
$N_{k}(t)\times\{k-1\}\sqsubseteq N_{l}(t)\times\{l-1\}$) and
$\tau-\{k\}\sqsubseteq\tau'-\{l\}$. Finally
$\tau_{\tup{k,t}}\sqsubseteq\tau'_{\tup{l,t}}$.
\end{proof}

\begin{lemma}[Monotony w.r.t.\ dominance]
\label{lem-mono-domin}
If $\tau\sqsubseteq \tau'$ then $\ubound_\tau(t)\leq \ubound_{\tau'}(t)$.
\end{lemma}
\begin{proof}
By induction over $\tau$. The base case, $\tau=\emptyset$, is covered with
$\ubound_\emptyset(t)=0$. For the inductive case, we assume that
$\tau\not=\emptyset$ so that $\ubound_\tau(t)$ is $1+\ubound_{\tau_{\tup{k,t}}}(t+1)$ for
some $k\in\tau$. With \autoref{lem-tau-k-t}.\textit{\ref{lem-tau-k-t-5}}, we
pick an $l\geq k$ such that
$\tau_{\tup{k,t}}\sqsubseteq\tau'_{\tup{l,t}}$. Then
\begin{align}
\ubound_\tau(t)
& = 1+\ubound_{\tau_{\tup{k,t}}}(t+1)
\tag{by assumption}
\\
&\leq 1+\ubound_{\tau'_{\tup{l.t}}}(t+1)
\tag{by ind.\ hyp., using \autoref{lem-tau-k-t}.\textit{\ref{lem-tau-k-t-5}}}
\\
& \leq \ubound_{\tau'}(t)\;.
\tag{by Eq.~\eqref{eq-def-Mtau}, since $l\in\tau'$}
\end{align}
\end{proof}

\lemmin*%
\begin{proof}
By induction over $\tau$. The case where $l=k$ is obvious so we assume
$l>k$ and hence $\{k,l\}\subseteq \tau$.
Now
\begin{align}
\ubound_{\tau_{\tup{k,t}}}(t+1)
& \geq 1+\ubound_{\tau_{\tup{k,t}\tup{l,t+1}}}(t+2)
\tag{by Eq.~\eqref{eq-def-Mtau}, since $l\in\tau_{\tup{k,t}}$}
\\
& \geq 1+\ubound_{\tau_{\tup{l,t}\tup{k,t+1}}}(t+2)
\tag{combining lemmata~\ref{lem-tau-k-t}.\textit{\ref{lem-tau-k-t-4}}
  and~\ref{lem-mono-domin}}
\\
& = \ubound_{\tau_{\tup{l,t}}}(t+1)\;.
\tag{by ind.\ hyp., since $k=\min\tau_{\tup{l,t}}$}
\end{align}
\end{proof}

Let us close this section on $\ubound$ with a consequence of
\autoref{lem-ubound-min}:
\begin{corollary}\label{cor-ubound-sum}
  Let $\tau=\emptyset$ or $\tau'\leq_m\{\min\tau\}$.  Then for all
  $t\geq 0$,
  \begin{equation*}
    \ubound_{\tau+\tau'}(t)=\ubound_{\tau'}(t)+\ubound_\tau(t+\ubound_{\tau'}(t))\;.
  \end{equation*}
\end{corollary}
\begin{proof}
  The statement is immediate if $\tau=\emptyset$.  Otherwise, we prove
  it by induction over $\tau'$.  The base case, $\tau'=\emptyset$, is
  covered with 
  \begin{equation*}
    \ubound_\tau(t)=
    0+\ubound_\tau(t+0)=\ubound_\emptyset(t)+\ubound_\tau(t+\ubound_\emptyset(t))\;.
  \end{equation*}
  For the inductive case, we assume $\tau'\neq\emptyset$, so that
  $k=\min\tau'$ exists and is no greater than $\min\tau$.  Then by
  \autoref{lem-tau-k-t}.\textit{\ref{lem-tau-k-t-1}},
  $\tau'_{\tup{k,t}}<_m\tau'$, and furthermore
  $\tau'_{\tup{k,t}}\leq_m\{\min\tau\}$.  Thus
  \begin{align*}
    \ubound_{\tau+\tau'}&(t)\\
    &=1+\ubound_{\tau+\tau'_{\tup{k,t}}}(t+1)\tag{by
      \autoref{lem-ubound-min}}\\
    &=1+\ubound_{\tau'_{\tup{k,t}}}(t+1)+\ubound_\tau(t+1+\ubound_{\tau'_{\tup{k,t}}}(t+1))\tag{by ind.\ hyp.}\\
    &=\ubound_{\tau'}(t)+\ubound_\tau(t+\ubound_{\tau'}(t))\;.\tag{by
      \autoref{lem-ubound-min}}
   \end{align*}
\end{proof}

\subsection{Proof of Proposition~\ref{prop:upper}}\label{ax:ubound}
\propupper*%
\begin{proof}
  We define in the next paragraph another function $\urec{k}$, which
  is monotone and such that $\urec{k}(x)\geq x$
  (\autoref{lem:monoG}).  It further belongs to $\FGH{\gamma+k-1}$
  by \autoref{lem:ihG}, and is such that
  $\uboundd{k}{r}(x)=\urec{k}^r(x)\cutsub x$ according to
  \autoref{lem:gG}, i.e.\ $\uboundd{k}{r}$ is defined through finite
  substitution from $\urec{k}$ and cut-off
  subtraction,\footnote{Cut-off subtraction
\begin{equation*}x\dot{-}y\eqdef\begin{cases}x-y&\text{if }x\geq
 y\\0&\text{otherwise},\end{cases}\end{equation*} can be defined by
 limited primitive recursion in $\FGH{0}$.} and therefore
  also belongs to $\FGH{\gamma+k-1}$.
\end{proof}

\paragraph{More about the Fast Growing Hierarchy}
Let us first give a few more details on the Fast Growing Hierarchy.
The class of functions $\FGH{k}$ is the closure of $\{\lambda
x.0,\lambda xy.x+y,\lambda x.x_i\}\cup\{F_n\mid n\leq k\}$ under the
operations of
\begin{description}
\item[substitution] if $h_0,h_1,\dots,h_n$ belong to the class, then
so does $f$ if
\begin{equation*}
  f(x_1,\dots,x_n)=h_0(h_1(x_1,\dots,x_n),\dots,h_n(x_1,\dots,x_n))
\end{equation*}
\item[limited recursion] if $h_1$, $h_2$, and $h_3$ belong to the
  class, then so does $f$ if
\begin{align*}
  f(0,x_1,\dots,x_n)&=h_1(x_1,\dots,x_n)\\
  f(y+1,x_1,\dots,x_n)&=h_2(y,x_1,\dots,x_n,f(y,x_1,\dots,x_n))\\
  f(y,x_1,\dots,x_n)&\leq h_3(y,x_1,\dots,x_n)\;.
\end{align*}
\end{description}
Here are a few monotonicity properties that will be useful in the following:
\begin{itemize}
\item for each $\alpha$ and all $n,x,y$ with $x>y$,
$F_\alpha^n(x)>F_\alpha^n(y)$ \citep[Lemma 2.6.(iii)]{fast},
\item for each $\alpha$ and all $m,n,x$, if $m>n$,
$F_\alpha^m(x)>F_\alpha^n(x)$ \citep[Lemma 2.6.(iv)]{fast}, and
\item for each $\alpha$ and every $k\geq 1$ we have, for all $n$ and
$x$, $F_{\alpha+k}^n(x)\geq F_\alpha^n(x)$ \citep[Lemma 2.8]{fast}%
.
\end{itemize}
\medskip

\paragraph{A Simpler Version of {$\ubound$}}
We consider a fast iteration hierarchy for $\uboundd{k}{r}$, given a monotone
unary function $f$:
\begin{align}
  \urec{1}(x)&\eqdef f(x)+x\label{eq:G0}\\
  \urec{k+1}(x)&\eqdef \urec{k}^{N_{k+1}(x)}(x+1)\;.\label{eq:Gn}
\end{align}

\begin{lemma}\label{lem:monoG}
  Let $f$ be a monotone unary function such that $f(x)\geq x$ and
  let $n\geq 1$.  Then the function $\urec{n}$ is monotone and such that
  $\urec{n}(x)\geq x$.
\end{lemma}

We leave the previous proof to the reader, and turn to the main
motivation for introducing $\urec{k}$:
\begin{lemma}\label{lem:gG}
  Let $k\geq 1$.  Then for all $r\geq 1$ and $x\geq 0$,
  \begin{equation*}
    \uboundd{k}{r}(x)=\urec{k}^r(x)-x\;.
  \end{equation*}
\end{lemma}
\begin{proof}
  We proceed by induction on types $\tau$ of form $r\times\{k\}$.  For the
  base case, which is $\tau=\{1\}$, we have for all $x$
  \begin{equation*}
    \uboundd{1}{1}(x)=f(x)=\urec{1}(x)-x\;.\tag{by Def.\ \eqref{eq:G0}}
  \end{equation*}
  For the induction step, we first consider the case $\tau=\{k\}$.
  Then, for all $x$,
  \begin{align}
   \uboundd{k}{1}(x)&=1+\uboundd{k-1}{N_{k}(x)}(x+1)\tag{by
   \autoref{lem-ubound-min}}\\
  &=1+\urec{k-1}^{N_{k}(x)}(x+1)-x-1\tag{by ind.\ hyp.}\\
  &=\urec{k}(x)-x\;.\tag{by Def.\ \eqref{eq:Gn}}
  \end{align}
  Finally, for the case $\tau=(r+1)\times\{k\}$, for all $x$,
  \begin{align*}
  &\!\!\!\!\uboundd{k}{r+1}(x)\notag\\
  &=1+\ubound_{r\times\{k\}+N_{k}(x)\times\{k-1\},f}(x+1)\tag{by
  \autoref{lem-ubound-min}}\\
  &=1+\uboundd{k-1}{N_{k}(x)}(x+1)\notag\\&\phantom{=}\:+\uboundd{k}{r}\!\left(x+1+\uboundd{k-1}{N_{k}(x)}(x+1)\right)\tag{by
  \autoref{cor-ubound-sum}}\\
  &=1+\urec{k-1}^{N_{k}(x)}(x+1)-x-1\notag{}\\
  &\phantom{=}\:+\uboundd{k}{r}\!\left(x+1+\urec{k-1}^{N_{k}(x)}(x+1)-x-1\right)\tag{by
  ind.\ hyp.\ on $\uboundd{k-1}{N_k(x)}$}\\
  &=\urec{k-1}^{N_{k}(x)}(x+1)-x+\uboundd{k}{r}\!\left(\urec{k-1}^{N_{k}(x)}(x+1)\right)\\
  &=\urec{k-1}^{N_{k}(x)}(x+1)-x+\urec{k}^r\!\left(\urec{k-1}^{N_{k}(x)}(x+1)\right)\notag\\&\phantom{=}\:-\urec{k-1}^{N_{k}(x)}(x+1)\tag{by
  ind.\ hyp.\ on $\uboundd{k}{r}$}\\
  &=\urec{k}^{r+1}(x)-x\;.\tag{by Def.\ \eqref{eq:Gn}}
\end{align*}
\end{proof}
  
\paragraph{Placing {$\urec{n}$} in the Fast Growing Hierarchy}
We prove the following lemma:
\begin{lemma}\label{lem:ihG}
  Let $\gamma\geq 1$ be an ordinal and $f$ be a unary monotone
  function in $\FGH{\gamma}$ with $f(x)\geq\max(1,x)$ for all $x$.  Then for
  all $k\geq 1$, $\urec{k}$ belongs to $\FGH{\gamma+k-1}$.
\end{lemma}
\begin{proof}
  Since $\gamma\geq 1$, and because $f$ is in $\FGH{\gamma}$, the
  function 
  \begin{equation}
  h(x)\eqdef k\cdot f(x)+x+1\;,
  \end{equation}
   defined through finite substitution from $f$ and addition, is
  monotone and also belongs to $\FGH{\gamma}$. Then, there exists
  $p\in\mathbb{N}$ such that, for all
  $x$ \citep[Theorem~2.10]{fast}:\footnote{The theorem is actually
  stated for a different version of $F_1$, but it turns out to hold
  with ours as well.}
  \begin{equation}\label{eq:finFk}
    h(x) < F_\gamma^p(x)\;.
  \end{equation}

  We start the proof of the lemma by several inequalities in
  Claims~\ref{cl:hpF} and~\ref{cl:majG}.
  \begin{claim}\label{cl:hpF}
    For all $y\geq 1$, and $x,n\geq 0$
    \begin{equation*}
      F_{\gamma+n}^{y\cdot h(x)}(x+1)\leq F_{\gamma+n+1}^{y\cdot(p+1)}(x)\;.
    \end{equation*}
  \end{claim}
  \begin{proof}
    We proceed by induction on $y$ for the proof of the claim.  If
    $y=1$, then
    \begin{align*}
    F_{\gamma+n}^{h(x)}(x+1)&\leq
  F_{\gamma+n}^{h(x)}\!\left(h(x)\right)\tag{since $h(x)\geq x+1$}\\
    &\leq F_{\gamma+n}^{h(x)+1}\left(h(x)\right)\tag{by monotonicity
  of $F_{\gamma+n}$}\\
    &=F_{\gamma+n+1}(h(x))
        \tag{by Def.\ \eqref{eq:Fn}}\\
    &<F_{\gamma+n+1}\left(F_\gamma^p(x)\right)\tag{by \eqref{eq:finFk} and monotonicity of $F_{\gamma+n+1}$}\\
    &\leq F_{\gamma+n+1}^{p+1}(x)\tag{by \citep[Lemma~2.8]{fast}}
  \end{align*}
   and the claim holds.  Quite similarly for the induction step,
    \begin{align*}
      \!\!F_{\gamma+n}^{(y+1)\cdot h(x)}\!(x+1)
      &=    F_{\gamma+n}^{h(x)}\!\left(F_{\gamma+n}^{y\cdot h(x)}(x+1)\right)\\
      &\leq F_{\gamma+n}^{h(x)}\!\left(F_{\gamma+n+1}^{y\cdot(p+1)}(x)\right)
        \tag{by ind.\ hyp.\ and monotonicity of $F_{\gamma+n}^{h(x)}$}\\
      &\leq F_{\gamma+n}^{h(x)}\!\left(h(F_{\gamma+n+1}^{y\cdot(p+1)}(x))\right)
        \tag{since $h(x)\geq x$ and by monotonicity of $F_{\gamma+n}^{h(x)}$}\\
      &\leq F_{\gamma+n}^{h(F_{\gamma+n+1}^{y\cdot(p+1)}(x))+1}\!\left(h(F_{\gamma+n+1}^{y\cdot(p+1)}(x))\right)
        \tag{since $F_{\gamma+n+1}^{y\cdot(p+1)}(x)\geq x$ and by
          monotonicity of $h$ and $F_{\gamma+n}^{y}(x)$}\\
      &=    F_{\gamma+n+1}\!\left(h(F_{\gamma+n+1}^{y\cdot(p+1)}(x))\right)
        \tag{by Def.\ \eqref{eq:Fn}}\\
      &<    F_{\gamma+n+1}\!\left(F_\gamma^p(F_{\gamma+n+1}^{y\cdot(p+1)}(x))\right)
        \tag{by \eqref{eq:finFk} and monotonicity of $F_{\gamma+n+1}$}\\
      &\leq F_{\gamma+n+1}^{p+1}\!\left(F_{\gamma+n+1}^{y\cdot(p+1)}(x)\right)
        \tag{by \citep[Lemma~2.8]{fast}}\\
      &=    F_{\gamma+n+1}^{(y+1)\cdot(p+1)}(x)\;.\tag*{\qedhere}
    \end{align*}
  \end{proof}

  \begin{claim}\label{cl:majG}
    For all $1\leq n\leq k$ and $x,y\geq 0$,
    \begin{equation*}
      \urec{n}^y(x)\leq F_{\gamma+n-1}^{y\cdot{(p+1)}^{n}}(x)\;.
    \end{equation*}
  \end{claim}
  \begin{proof}
    Let us first show that, for all $1\leq n\leq k$,
    \begin{equation}\label{eq:GyFy}
    \forall x.\urec{n}(x)\leq F_{\gamma+n-1}^{{(p+1)}^{n}}\!(x)\text{
      implies }\forall x,y.\urec{n}^y(x)\leq F_{\gamma+n-1}^{y\cdot{(p+1)}^{n}}\!\!(x).
    \end{equation}
    By induction on $y$: for $y=0$,
    $\urec{n}^0(x)=x=F_{\gamma+n-1}^{0\cdot{(p+1)}^{n}}(x)$, and for the
    induction step on $y$, for any $x$, $y$,
    \begin{align*}
      \urec{n}^{y+1}(x)&=\urec{n}\!\left(\urec{n}^y(x)\right)\\
      &\leq \urec{n}\!\left(F_{\gamma+n-1}^{y\cdot{(p+1)}^{n}}\!(x)\right)
        \tag{by ind.\ hyp.\ and monotonicity of $\urec{n}$}\\
      &\leq F_{\gamma+n-1}^{{(p+1)}^{n}}\!\left(F_{\gamma+n-1}^{y\cdot{(p+1)}^{n}}\!(x)\right)
        \tag{by ind.\ hyp.}\\
      &= F_{\gamma+n-1}^{(y+1)\cdot{(p+1)}^{n}}(x)\;.
    \end{align*}

    It remains to prove that $\urec{n}(x)\leq
    F_{\gamma+n-1}^{{(p+1)}^n}\!(x)$ by induction on $n$: for $n=1$,
    \begin{equation*}
    \urec{1}(x)=f(x)+x\leq h(x)<F^p_\gamma(x)\leq F^{p+1}_{\gamma+1-1}(x)
    \end{equation*}
    by \eqref{eq:finFk} and monotonicity of $F_\gamma$.  For the induction
    step on $n$,%
    \begin{align*}
      \urec{n+1}(x)&=\urec{n}^{N_{n+1}(x)}\!(x+1)\\
      &\leq F_{\gamma+n-1}^{N_{n+1}(x)\cdot{(p+1)}^{n}}\!(x+1)
        \tag{by ind.\ hyp.\ and \eqref{eq:GyFy} for $y=N_{n+1}(x)$}\\
      &\leq F_{\gamma+n-1}^{h(x)\cdot{(p+1)}^{n}}\!(x+1)\tag{since
    $n\leq k$}\\
      &\leq F_{\gamma+n}^{(p+1)\cdot{(p+1)}^{n}}\!(x)\tag{by
        \autoref{cl:hpF} for $y=(p+1)^{n}\geq 1$}\\
      &=F_{\gamma+n}^{{(p+1)}^{n+1}}(x)\;.\tag*{\qedhere}
    \end{align*}
  \end{proof}

  The main proof consists in first proving that for all $1\leq n\leq k$,
  \begin{equation}\label{eq:Gy-in-F}
    \lambda x.\urec{n}(x)\in\FGH{\gamma+n-1}\text{ implies
    }\lambda xy.\urec{n}^y(x)\in\FGH{\gamma+n}\;.
  \end{equation}
  Indeed, for all $x$, $y$,
  \begin{align*}
    \urec{n}^y(x)
    &\leq F_{\gamma+n-1}^{y\cdot {(p+1)}^{n}}(x)\tag{by \autoref{cl:majG}}\\
    &\leq F_{\gamma+n-1}^{x+y\cdot {(p+1)}^{n}+1}\!\left(x+y\cdot {(p+1)}^{n}\right)\tag{by monotonicity of $F_{\gamma+n-1}$}\\
    &= F_{\gamma+n}\!\left(x+y\cdot {(p+1)}^{n}\right)\;.
  \end{align*}
  Thus $\lambda xy.\urec{n}^{y}(x)$ is defined by a simple recursive definition
  from $\urec{n}$, which is in
  $\FGH{\gamma+n-1}\subseteq\FGH{\gamma+n}$ by hypothesis, and
  is limited by a function in $\FGH{\gamma+n}$, namely $\lambda
  xy.F_{\gamma+n}(x+y\cdot {(p+1)}^{n})$, clearly defined by finite
  substitution from addition and $F_{\gamma+n}$.  It belongs therefore to
  $\FGH{\gamma+n}$.

  It remains to prove that for all $1\leq n\leq k$, $\urec{n}$ is in
  $\FGH{\gamma+n-1}$.  We proceed by induction on $n$; for the case
  $n=1$, $\urec{1}=f(x)+x$ is defined by finite substitution from $f$
  and addition, thus belongs to $\FGH{\gamma}$ by hypothesis.
  For the induction step on $n$, $\lambda x.\urec{n+1}(x)=\lambda
  x.\urec{n}^{N_{n+1}(x)}(x+1)$ is defined by substitution from
  \begin{itemize}
  \item addition,
  \item $\lambda x.N_{n+1}(x)=\lambda x.(n+1)\cdot(f(x)\cutsub 1)$,
  which is defined through cut-off subtraction (recall that
  $f(x)\geq 1$ for all $x$), $f$, and addition, and thus belongs to
  $\FGH{\gamma}\subseteq\FGH{\gamma+n}$, and from\
  \item $\lambda xy.\urec{n}^y(x)$, which is by induction hypothesis
  and Eq.~\eqref{eq:Gy-in-F} in $\FGH{\gamma+n}$.
  \end{itemize}                
  Thus $\lambda x.\urec{n+1}(x)$ belongs to $\FGH{\gamma+n}$.
\end{proof}

\subsection{Proof of Proposition~\ref{prop-llex-fgh}}\label{ax:lbound}
\proplower*%
\begin{proof}
  Let us first show that for all $k\geq 1$
  \begin{multline}\label{eq:GnqFmq}%
    \forall t\,.\,\llexd[F_\gamma]{k}(t)\geq F_{\gamma+k-1}(t)\text{ implies
      }\forall r\geq 1,t\,.\,\llexd[F_\gamma]{k}^r(t)\geq F_{\gamma+k-1}^r(t)\;.
  \end{multline}
  By induction on $r$; the base case for $r=1$ holds by hypothesis,
  and the induction step holds by monotonicity of
  $F_{\gamma+k-1}$.

  It remains to prove $\llexd[F_\gamma]{k}(t)\geq F_{\gamma+k-1}(t)$ by
  induction over $k\geq 1$.  The base case is settled by
  $\llexd[F_\gamma]{1}(t)=F_\gamma(t)=F_{\gamma+1-1}(t)$, and for the induction
  step, we have for all $t\geq 0$:
  \begin{align*}
    \llexd[F_\gamma]{k+1}(t)&=\sum_{j=1}^{F_\gamma(t)}\lbound[F_\gamma]{k}\!\left(\Lbound[F_\gamma]{k}^{j-1}(t)\right)\\
    &\geq \sum_{j=1}^{F_\gamma(t)}\lbound[F_\gamma]{k}^j(t)\tag{by
      monotonicity of $\llex$}\\
    &\geq\lbound[F_\gamma]{k}^{F_\gamma(t)}(t)\tag{still by monotonicity of $\llex$}\\
    &\geq F^{F_\gamma(t)}_{\gamma+k-1}(t)\tag{by ind.\ hyp.\ and
      \eqref{eq:GnqFmq}}\\
    &\geq F^{t+1}_{\gamma+k-1}(t)\tag{by monotonicity of
      $F_{\gamma+k-1}$}\\
    &=F_{\gamma+k}%
  \end{align*}

  Finally, for all $r\geq 1$ and $t\geq 0$,
  \begin{align*}
    \llexd[F_\gamma]{r,k}(t)&\geq\sum_{j=1}^r\lbound[F_\gamma]{k}\!\left(\Lbound[F_\gamma]{k}^{j-1}(t)\right)\\
    &\geq\sum_{j=1}^r\llexd[F_\gamma]{k}^j(t)\tag{by monotonicity of $\llex$}\\
    &\geq\llexd[F_\gamma]{k}^r(t)\tag{still by monotonicity of $\llex$}\\
    &\geq F_{\gamma+k-1}^r(t)\;.\tag{by \eqref{eq:GnqFmq} and the
  previous argument}
  \end{align*}
\end{proof}